\documentstyle[aaspp]{article}
\def\kms {km~s$^{-1}$}

\begin{document}
\title{New Protostellar Collapse Candidates: An HCO$^{+}$ Survey of the Class 0
Sources}
\author {Erik M. Gregersen\altaffilmark{1} and Neal J. Evans
II\altaffilmark{2}}
\affil{Department of Astronomy, The University of Texas at Austin,
       Austin, TX 78712--1083}
\author{Shudong Zhou\altaffilmark{3}}
\affil{Department of Astronomy, University of Illinois, Urbana, IL 61801}
\author{Minho Choi\altaffilmark{4}}
\affil{Institute of Astronomy and Astrophysics, Academia Sinica, P. O.
Box 1-87, Nankang, Taiwan 115, R.O.C.}
\altaffiltext{1}{Electronic mail: erik@astro.as.utexas.edu}
\altaffiltext{2}{Electronic mail: nje@astro.as.utexas.edu}
\altaffiltext{3}{Electronic mail: zhou@astro.uiuc.edu}
\altaffiltext{4}{Electronic mail: minho@biaa3.biaa.sinica.edu.tw}
\centerline{\footnotesize {\LaTeX}ed at \number\time\ min., \today}

\begin{abstract}

We have observed 23 Class 0 sources in the HCO$^{+}$ $J=4-3$ and $3-2$ lines.
The mean bolometric temperature of the 16 sources with well-determined values
is 44 K and the mean luminosity is 5.7 L$_{\sun}$, excluding two sources of
considerably higher luminosity.  Nine sources, including three sources
previously suggested to be collapsing, have the correct (blue) spectral line
asymmetry for infall in both lines.
Three sources have the opposite (red) asymmetry in both lines, and one source,
L1157, has a red asymmetry in HCO$^{+}$
$J=4-3$ and a blue asymmetry in $J=3-2$.  The rest have no
significant or consistent asymmetry.  The H$^{13}$CO$^{+}$ $J=4-3$ and
$3-2$ lines were also observed to find the velocity of the ambient gas, and
sources with an interesting line asymmetry were mapped.  A Monte Carlo code was
used to produce an
evolutionary sequence of collapsing cloud models of the HCO$^{+}$ $J=4-3$ and
$3-2$ lines and
to compare various diagnostics of the resulting line profiles. The same
code was used
to compare infall models to the observations in one source, L1527. The
results were
consistent with previous collapse models. Based on integrated intensity
maps
of the line peaks and wings,
as well as the velocity of the H$^{13}$CO$^{+}$ line,
we select six of the nine sources with a blue line asymmetry
as good {\it candidates} for protostellar collapse.
Further evidence is needed to establish that infall is taking place.
The HCO$^+$ spectra are not conclusive because bipolar outflows produce strong
emission, which can confuse the issue in any individual source.
However, the predominance of blue asymmetries over red asymmetries is
not naturally
explained in outflow models, whereas it is expected in collapse models.

\end{abstract}


\section {INTRODUCTION}

	An important aim of the study of young stellar objects (YSOs)
has been to
classify them in an evolutionary sequence and identify when in this
sequence protostellar
collapse occurs.  The most widely used classification scheme is that of
Lada and Wilking (1984) and Lada
(1987) which divides YSOs into 3 categories based on the slope of their
 infrared spectral energy distribution (SED).  The oldest population, Class
III, have a blackbody-like SED and are interpreted as pre-main-sequence stars
without a disk.  Class II sources have
a flat or negatively-sloped SED in the infrared
and are
understood as classical T Tauri stars with an optically thick disk
of circumstellar dust.
Class I sources have an SED that rises with wavelength up to 100 $\mu$m and are
believed to be collapsing protostars (Adams et al.\ 1987; Wilking et al.\
1989a).

	Recently, Andr\'e, Ward-Thompson, and Barsony (1993) (hereafter AWB)
noted that several sources have not been detected in the
near-infrared and have blackbody-like SEDs that peak
in the submillimeter,
suggesting extremely low dust temperatures ($T_{D}\sim$ 20 K).  They
called these
sources Class 0 (which they defined as a submillimeter source with
L$_{bol}$/L$_{1.3 mm}$ $\lesssim$ 2 $\times$ 10$^{4}$)
because their cold temperature suggested an object even younger than typical
Class I objects. Barsony (1994) listed the following as characteristics of
these sources in addition to the low L$_{bol}$/L$_{1.3 mm}$: a SED like a
30 K blackbody, undetected at
$\lambda <$ 10 $\mu$m, and a molecular outflow. Chen et al.\ (1995) found
that Class 0 sources have bolometric temperatures (T$_{bol}$) less than 70 K.
T$_{bol}$ is the temperature of a blackbody with the same mean frequency as
the SED of the source (Myers \& Ladd 1993).

Andr\'e and Montmerle (1994) argued that Class 0 sources
have more circumstellar mass than stellar mass.  They suggested that Class
0 sources are
still in their active accretion phase, while Class I sources
have already accreted most of their mass. If this suggestion is correct,
Class 0 sources should be good candidates to show kinematic evidence of
collapse.

	The goal of this project was to survey all of the Class 0 objects that
had been cited in the astronomical literature for signs of protostellar
collapse.  Since we were unable to determine if every claimed Class 0
object meets the
criterion of AWB or Barsony, the selected sample of 23 sources
(see Table 1) includes any source that has been
identified as either a definite or a probable Class 0 source.
Although this criterion is rather loose, we have checked the credentials
of our sources by computing T$_{bol}$, using the procedure of Myers \& Ladd
(1993), for all sources with sufficient data. Since many of these sources
have few spectral data points, we relaxed one criterion of Myers \& Ladd.
They required six spectral data points but we used as few as four, which
seemed to be adequate as long as the photometric data lay on either side of
the peak of the SED.
The values of T$_{bol}$ are given in Table 2, along with information on
L$_{bol}$, distance, and references.  All 16 of the sources for which
T$_{bol}$ can be calculated lie in the Class 0 category
(T$_{bol} < 70$ K) within error
bars. The mean T$_{bol}$ is 44 K and the mean L$_{bol}$ is 5.7
L$_{\sun}$, excluding
the extremely luminous sources IRAS 20050 and SMM1. This sample is primarily
one of very
embedded, low luminosity objects.
For two other sources, there are enough photometric data
to calculate upper limits (if data are available on the blue side of the peak)
or lower limits (if data are available on the red side of the peak) for
T$_{bol}$.  Some of the sources without sufficient data to compute
T$_{bol}$ or L$_{bol}$ lie in regions of considerably higher luminosity, where
confusion prevents separation of the contributions of different sources in the
far-infrared.

The method of spectral line asymmetry (Leung \& Brown 1977; Zhou 1992)
was used to identify collapse candidates.
An optically thick line in a collapsing core shows a blue
asymmetry in the line profile toward the collapse center,
while an optically thin line appears symmetric.  The
optically thick lines we used were the
$J=4-3$ and $3-2$ transitions of HCO$^{+}$, and the optically thin lines were
those of rarer isotopomers of HCO$^{+}$, like H$^{13}$CO$^{+}$ and
HC$^{18}$O$^{+}$.  Clumpy outflows can also
produce peaks that may be either red or blue; we address the possible
confusion between infall and outflow in \S 4.3.

\section {OBSERVATIONS AND RESULTS}

Observations were made of twenty-three Class 0 sources at the 10.4-m
telescope of the Caltech Submillimeter Observatory (CSO)\footnote{The CSO is
operated by the California Institute of Technology under funding from the
National Science Foundation, contract AST 90--15755.} at Mauna Kea, Hawaii in
1994 October, 1994 November, 1994 December, 1995 March, 1995 June, 1995
December, and 1996 June.
The sources are listed in Table 1 with their celestial coordinates
and the off position used for position switching.  The off positions were
checked for absence of detectable HCO$^{+}$ emission.  We used SIS receivers
operating
at 230 and 345 GHz, and the backend was a 1024-channel, acousto-optic
spectrometer
with a bandwidth of 49.5 MHz.
The frequency resolution varied over the five runs from about 2.5 to 3
channels, which corresponds to a velocity resolution of 0.14 to 0.17 \kms\ at
260 GHz and
of 0.10 to 0.12 \kms\ at 357 GHz.  Chopper-wheel calibration was used to obtain
the antenna
temperature, $T_A^*$.  The lines we observed and their frequencies are listed
in
Table 3 along with the main beam efficiency, the velocity resolution, and the
beamsize at
each frequency.  The efficiencies were calculated from numerous observations of
planets.
Data from separate runs were resampled to the resolution of the run with
the worst frequency resolution before averaging.  A first order baseline was
removed
before scans were averaged.  Seven sources which had
noticeable asymmetry (either to the blue or red side) at the central position
were mapped (map sizes ranged from 30\arcsec\ by 30\arcsec\ to 60\arcsec\ by
90\arcsec) and observed in at least one transition of H$^{13}$CO$^{+}$.

Twenty-three sources were observed in this survey. All twenty-three sources
were observed in the HCO$^{+}$ $J=3-2$ line.  Eighteen sources were observed in
the HCO$^{+}$ $J=4-3$ line.  Nineteen and four sources were observed in the
H$^{13}$CO$^+$ $J=3-2$ and $4-3$ lines, respectively.  Seven sources were
observed in the $J=3-2$ line of HC$^{18}$O$^+$.

Nine sources showed blue asymmetry in both the the HCO$^{+}$ $J=4-3$ and $3-2$
lines.  Three sources had a red
asymmetry in both lines and one source, L1157, had opposite asymmetry in each
of the two lines.  Five sources showed no asymmetry in both lines.  Five other
sources were observed only in the $J=3-2$ line.  Of the five sources observed
only in the $J=3-2$ line, three
showed no asymmetry, one showed a red asymmetry, and one showed a blue
asymmetry.

The line properties are listed in Table 4.  $T_A^*$ is the
highest temperature in the line profile.  For singly-peaked lines, the
velocity and $\Delta$V are the centroid and width of the line,
found by fitting a single Gaussian to the line profile.
For lines that have two peaks, we list two
values of $T_A^*$ and velocity, one for each peak, and we give one value for
the line width, which is the full width across the spectrum at the temperature
where the weaker peak falls to half power.

\section {Individual Sources}

\subsection {L1448-N}

This core in Perseus ($d$ = 300 pc) is coincident with IRS 3 and lies in the
outflow lobe of L1448-C.  An
H$_{2}$O maser coincides with the peak of the ammonia emission (Anglada et al.\
1989).
Bachiller et al.\ (1990) mapped the outflow in CO and found it to be quite weak
and not
well collimated.  A clear red asymmetry is seen in HCO$^{+}$ $J=3-2$
(Figure 5).

\subsection {L1448-C}

Bachiller et al.\ (1990) discovered this highly collimated outflow ($d$ = 300
pc) which had no known
exciting source.  Curiel et al.\ (1990) found a radio continuum source at the
center of the
outflow.  Line asymmetries consistent with infall have been seen in the 85 GHz
line of
C$_{3}$H$_{2}$ (Mardones et al.\ 1994).  The HCO$^{+}$ $J=4-3$ and $3-2$ lines
display an asymmetry with the red peak stronger than the blue peak. The
H$^{13}$CO$^{+}$ $J=3-2$ line peaks on the red side of the self-absorption dip,
but the
H$^{13}$CO$^{+}$ $J=4-3$ line peaks on the red peak of the HCO$^{+}$ $J=4-3$
line
(Figures 3 and 4), suggesting that the blue peak is an optically thin clump of
outflowing gas. In support of this interpretation, we note that the velocity of
the blue peak in the HCO$^+$ spectrum is well outside the line profile seen in
C$_{3}$H$_{2}$ by Mardones et al.  Consequently, our data do not provide
evidence for or against the suggestion by Mardones et al.\ that L1448-C may be
collapsing.

\subsection {NGC 1333 IRAS 4A}

Jennings et al.\ (1987) discovered NGC 1333 IRAS 4 ($d$ = 350 pc) in IRAS CPC
observations.
Submillimeter and millimeter photometry by Sandell et al.\ (1991) revealed a
double source
(IRAS 4A and 4B) with a 31\arcsec\ separation and a SED that peaks at about 500
$\mu$m.
Recent interferometer observations have shown that IRAS 4A is itself
a binary with a separation of 1\farcs8 (Lay et al.\ 1995).  Our
HCO$^{+}$ $J=4-3$ and $3-2$ data show a blue asymmetry, with the
H$^{13}$CO$^{+}$ $J=3-2$ line slightly offset
to the blue from the self-absorption dip (Figure 1).

\subsection {NGC 1333 IRAS 4B}

This source ($d$ = 350 pc) appears to be a system of more than two stars (Lay
et al.\ 1995).  A strong blue asymmetry is seen in the HCO$^{+}$ $J=3-2$ line
(Figure 1). The red peak
is very weak in the HCO$^{+}$ $J=4-3$ line (Figure 2).  The slight blue offset
of the
H$^{13}$CO$^{+}$ $J=3-2$ line from the absorption dip that was seen in IRAS 4A
is
also present here.

\subsection{L1551-NE}

This source ($d$ = 140 pc) lies in a lobe of the famous L1551 outflow.
Moriarty-Schieven et al.\ (1995a) identified this as an object between
Class I and Class 0; we find a T$_{bol}$ of 75$\pm$18 K.  CS $J=3-2$ and
$J=5-4$ observations show a blue
asymmetric line profile (Moriarty-Schieven et al.\ 1995b).  Our HCO$^{+}$
$J=3-2$ observations show a multi-peaked line profile (Figure 5). The blue peak
in our HCO$^{+}$ $J=3-2$ spectrum has the same velocity as that observed in the
double-peaked CS spectra of Moriarty-Schieven et al.\ (1995b).  The
H$^{13}$CO$^{+}$ $J=3-2$ line has two peaks and the HCO$^{+}$
$J=3-2$ has three.  The spectrum is too complex to decide whether collapse is
occurring.

\subsection {L1527}

The spectral energy distribution of this source ($d$ = 140 pc) is quite similar
to those
of B335 and L483 (Ladd et al.\ 1991).  Zhou et al.\ (1994), Mardones et al.\
(1994), and  Myers et al.\
(1995) have presented observations that show infall asymmetry in the H$_{2}$CO
$J=3_{12}-2_{11}$ line, the 85 GHz line of C$_{3}$H$_{2}$, and the H$_{2}$CO
$J=2_{12}-1_{11}$ line, respectively.  Our HCO$^{+}$ $J=4-3$ and $3-2$ data
also
show a prominent infall signature.  The H$^{13}$CO$^{+}$ $J=4-3$ and $J=3-2$
lines
peak in the self-absorption dips of the HCO$^{+}$ $J=4-3$ and $3-2$ line
profiles
(Figures 1 and 2).

\subsection {RNO43MM}

No emission has been detected from this source ($d$ = 400 pc) at 12 or 25
$\mu$m (IRAS PSC) and its
spectral energy distribution peaks at about 100 $\mu$m (Zinnecker et al.\
1992).
The derived $T_{bol}$ of 36 K confirms that it is a Class 0 source.
Our HCO$^{+}$ $J=3-2$ data show a symmetric line (Figure 5).

\subsection {NGC 2024 FIR 5}

These two sources (FIR 5 and FIR 6) ($d$ = 415 pc) were found by
Mezger et al.\ (1988) who
suggested that they were isothermal protostars.  A unipolar jet has been
seen near this source but the identification of FIR 5 as the driving source
is uncertain (Richer et al.\ 1992).  The identification of FIR 5 and 6 as
isothermal
protostellar condensations is controversial because warm gas is present
in these cores,
suggesting that the objects may be more evolved than Mezger et al.\ (1988)
believed (Lis et
al.\ 1991).  CS observations suggest that these two sources are massive enough
to form stars (Chandler and Carlstrom 1996).  No collapse asymmetry is seen in
the HCO$^{+}$ $J=3-2$ line
but a low
velocity shoulder was detected (Figure 5).

\subsection {NGC 2024 FIR 6}

A bipolar outflow with a very young dynamical time scale ($<$400 yr) is
present (Richer 1990).  An H$_{2}$O
maser is associated with this object ($d$ = 415 pc) (Johnston et al.\ 1973).
Barnes and Crutcher (1990) observed a torus around
this source, perpendicular to the the outflow
(see also Lis et al.\ 1991). Like NGC 2024
FIR 5, a low velocity shoulder was detected in the
HCO$^{+}$ $J=3-2$ line (Figure 5).

\subsection {HH25MMS}

This source ($d$ = 400 pc) is within the L1630 cloud.  Gibb and Heaton (1993)
detected
a CO outflow centered on this source.  Bontemps et al.\ (1995) identified this
source as
either a Class 0 or a Class I source.  The HCO$^{+}$ $J=4-3$ line at the
central
position has a clear blue asymmetry but
the HCO$^{+}$ $J=3-2$ line is only slightly asymmetric (Figures 1 and 2).

\subsection {IRAS 08076-3556}

This source ($d$ = 400 pc) is located in the Gum Nebula and is the exciting
star of HH120.  It has a steep
spectral index in the infrared (Persi et al.\ 1990).  Persi et al.\ (1994)
noted
that the 1.3 mm continuum flux was similar to those of other Class
0 sources.  Our calculated T$_{bol}$, 74$\pm$15 K, puts it on the boundary of
the Class 0 category.  Our
observations show a symmetric line in HCO$^{+}$ $J=4-3$ and and a slight blue
asymmetry in $J=3-2$  (Figures 5 and 6).

\subsection {VLA 1623}

This source ($d$ = 125 pc), the prototype for the Class 0 category, has a
highly collimated
CO outflow that extends over 15\arcmin\ in length (Andr\'e et al.\ 1990a; Dent
et al.\ 1995).
AWB (1993) argued that the low temperature, high extinction, and massive
envelope
of VLA 1623 were evidence of an extremely young object.
Our observations are ambiguous regarding collapse.
The HCO$^{+}$ $J=3-2$ and HCO$^{+}$ $J=4-
3$ lines both have a three-peaked spectrum with red asymmetry (Figures 3 and
4).  The H$^{13}$CO$^{+}$
$J=3-2$ line peaks between the middle and the red peak of the HCO$^{+}$ $J=3-2$
line (Figure 3).  These line profiles probably result from confusion with
clumps in the outflow.

\subsection {IRAS 16293--2422}

Walker et al.\ (1986) reported the first observations of spectral line
signatures of infall in this
source ($d$ = 125 pc).  Menten (1987) disputed this interpretation
and claimed that the asymmetry was a rotational effect.  However, Zhou (1995)
has
modeled this core as collapse with rotation.  There is an elongated structure
in the central
region which was originally attributed to a disk-like structure (Mundy et al.\
1986) but is
now known to be a binary system with a separation
of 5\arcsec\ (Wootten 1989).  The
HCO$^{+}$ $J=4-3$ and $J=3-2$ lines are quite strong (15 K and 18 K,
respectively)
and have a clear infall signature (Figures 1 and 2). Even the H$^{13}$CO$^{+}$
$J=3-2$
line shows central self-absorption (Figure 1) suggesting it is optically thick,
as van
Dishoeck et al.\ (1995) found for the $J=4-3$ line. Our detection of the
HC$^{18}$O$^{+}$ $J=3-2$ line (Figure 7) would agree with the interpretation of
van Dishoeck et al. The HC$^{18}$O$^{+}$ $J=3-2$ line profile may have two
peaks, suggesting
self-absorption.

\subsection {L483}

The CO outflow ($d$ = 200 pc) was observed by Parker et al.\ (1991) who noted
its similarity
to that of B335 in that it has some blue emission in the red lobe and vice
versa.
Fuller and
Myers (1993) traced the dense gas in the cloud with the HC$_{3}$N $J=4-3$ line.
An
H$_{2}$O maser has been detected by Wilking et al.\ (1994).  Fuller et al.\
(1995)
calculated a T$_{bol}$ of 46 K, similar to the 48$\pm$5 K we calculated.  Myers
et al.\
(1995) observed a double-peaked line profile with a stronger blue peak in the
H$_{2}$CO $J=3_{12}-2_{11}$ and $J=2_{12}-1_{11}$ lines.
The HCO$^{+}$ $J=3-2$ line profile has a red-peaked
spectral line profile with H$^{13}$CO$^{+}$ $J=3-2$ peaking to the red side of
the dip,
while the HCO$^{+}$ $J=4-3$ line displays a slight red asymmetry (Figures 3 and
4).
Unlike the case of L1448-C, the velocities of the red and blue peaks in HCO$^+$
agree with those in the H$_2$CO spectra of Myers et al.
The fact that HCO$^{+}$ has a red asymmetry weakens the case for collapse.

\subsection {Serpens SMM1}

This source in Serpens ($d$ = 310 pc) is also known as FIRS1 (Harvey et al.\
1984).  Casali et al.\ (1993)
detected this source,
discovered the next three sources in submillimeter and millimeter
continuum photometry of the Serpens molecular cloud core, and noted the lack of
near-IR
counterparts, a trait that many other Class 0 objects share.
However, they also pointed out that this
could be a more evolved object because a nearby jet shows reflected 2 $\mu$m
emission from a possible hot source at the center.  Hurt et al.\ (1996b)
observed  a double-peaked and symmetric line profile with a dip at 9 \kms\ in
the $J=3_{03}-2_{02}$ line of H$_{2}$CO.
The HCO$^{+}$ $J=4-3$ and $J=3-2$ lines reveal quite strong self-absorption of
a
symmetric line suggesting a large optical depth (Figures 5 and 6), but there is
no
collapse signature, in  agreement with Hurt et al.\ (1996b).

\subsection {Serpens SMM4}

This source ($d$ = 310 pc) and the following two sources were discovered in the
submillimeter
continuum maps of Casali et al.\ (1993).  Hurt et al.\ (1996b) saw a slight
scallop at 8.75 \kms\ in the $J=3_{03}-2_{02}$ line of H$_{2}$CO.
The HCO$^{+}$ $J=4-3$ and $J=3-2$ lines display a strong blue
asymmetry (Figures 1 and 2).  The H$^{13}$CO$^{+}$ $J=3-2$ line peaks in the
self-absorption dip of the HCO$^{+}$ $J=3-2$ line which agrees in velocity with
the dip in the 218 GHz H$_{2}$CO line.

\subsection {Serpens SMM2}

The 218 GHz H$_{2}$CO line shows two peaks (Hurt et al.\ 1996b).  A slight blue
asymmetry can be seen in the HCO$^{+}$ $J=3-2$ line at the central
position ($d$ = 310 pc).  The blue asymmetry is also present in the
HCO$^{+}$ $J=4-3$ line, but it is not as prominent (Figure 2).
 The H$^{13}$CO$^{+}$
$J=3-2$ line has two velocity components at the same velocity as the two peaks
of the
HCO$^{+}$ $J=3-2$ line (Figure 1), suggesting that two cloud components exist
in
this region, rather than collapse.   The HC$^{18}$O$^{+}$ $J=3-2$ line
(Figure 7) has one component, but higher signal-to-noise
observations are needed.

\subsection {Serpens SMM3}

A blue outflow wing can be seen in the HCO$^{+}$ $J=3-2$ line, but the central
line core ($d$ = 310 pc)
displays no asymmetry in either the HCO$^{+}$ $J=4-3$ or $J=3-2$ line (Figures
5 and
6).

\subsection {L723}

Davidson (1987) identified this far-infrared and submillimeter
source ($d$ = 300 pc).  The outflow of this cloud has a quadrupolar
structure (Goldsmith et al.\ 1984), which Avery et al.\ (1990) interpreted as
the
limb-brightened walls of a cavity.  Anglada et al.\
(1991) found two radio continuum sources within L723, suggesting two outflows,
but Cabrit and Andr\'e (1991) reported a single source. The Class 0
identification
(Cabrit and Andr\'e) is supported by our calculation of T$_{bol} = 39$ K.
Our HCO$^{+}$ observations show a symmetric line
(Figures 5 and 6) .

\subsection {B335}

B335 is a well-studied Class 0 source ($d$ = 250 pc).  Zhou et al.\
(1990) mapped the 6 cm H$_{2}$CO line in this source and found a density
gradient
consistent with the Shu (1977) model of inside-out collapse.  More direct
evidence of collapse was
found by Zhou et al.\ (1993) with the observation of an asymmetric spectral
line signature
in the CS $J=2-1$ and $J=3-2$ lines and the 140 and 225 GHz lines of H$_{2}$CO.
Choi et al.\ (1995) have modeled the line profiles of Zhou et al.\ (1993) as
inside-out
collapse.  Velusamy et al.\ (1996) have seen clumpy structure in the infalling
envelope.  Our central position is taken from the 2.7 mm continuum observations
of Chandler and Sargent (1993).  Our HCO$^{+}$ observations also show an infall
signature (Figure 1 and
2).

\subsection {IRAS 20050+2720}

This core ($d$ = 700 pc) was included in the sample of cold IRAS sources of
Wilking et al.\ (1989b).  The
value of 69$\pm 7$ for T$_{bol}$ places it barely within the category of Class
0
objects.  However, it is considerably more luminous than the other sources.
Recently, Bachiller et al.\ (1995) discovered an outflow with three pairs of
lobes
suggesting more than one central star. One outflow has an extremely high
velocity jet with
molecular bullets like L1448-C.  The CS $J=3-2$ and C$^{34}$S $J=3-2$ line
observations of Bachiller et al.\ (1995) show the optically thin C$^{34}$S
peaking at 6.1
\kms, quite close to the CS self-absorption dip at 6 \kms.  Chen et al.\ (1996)
have detected a cluster of 200 near-infrared sources.  The HCO$^{+}$ $J=3-2$
data show a broad double-peaked line toward the center.  The H$^{13}$CO$^{+}$
$J=3-2$ line peaks
in the absorption dip of the HCO$^{+}$ $J=3-2$ line but is rather broad (Figure
1).  The
HCO$^{+}$ $J=4-3$ line
shows a slight blue asymmetry toward the central position (Figure 2).

\subsection {S106FIR}

This source ($d$ = 600 pc) lies in the bipolar nebula S106.  S106 is bisected
by a dark lane that has possibly contains a disk (Bally et al.\ 1983).
However, more recent observations have shown that was what believed to be a
disk is actually several clumps (Barsony et al.\ 1989, Richer et al.\ 1993).
This source was first detected by Richer et al.\ in continuum maps at 450, 850,
and
1100 $\mu$m at a position 15\arcsec\ west of S106IR.  Its position coincides
with an
H$_2$O maser (Stutzki et al.\ 1982).  No asymmetry was detected, but broad line
wings,
possibly from an outflow, were seen in the HCO$^{+}$ $J=4-3$ and $J=3-2$ lines
(Figures 5 and 6).

\subsection {L1157}

This source ($d$ = 440 pc) has a well-collimated CO outflow in which the blue
lobe of the outflow seems
to be shock-heating the ambient gas  (Umemoto et al.\ 1992).  Tafalla and
Bachiller (1995)
pointed out that the SED is quite similar to that of a Class 0 object; we find
T$_{bol}
\lesssim 44$ K.
Our HCO$^{+}$ $J=3-2$ data shows the required blue asymmetry for collapse, but
the HCO$^{+}$ $J=4-3$ line shows the opposite asymmetry (Figure 3 and 4).
  The blue outflow wing is quite clear at the center position in HCO$^{+}$
$J=3-2$.  The H$^{13}$CO$^{+}$ $J=3-2$ line peaks in the self-absorption dip of
the HCO$^{+}$ $J=3-2$ line.

\section {DISCUSSION}

The line profiles in nine sources are consistent
with predictions of line profiles from collapsing clouds. Caution is suggested
by the fact that three sources show the opposite asymmetry and one shows
different
asymmetries in different lines. In this section, we will present some
predictions
for the expected shape of HCO$^+$ lines in collapsing clouds. We will then
use these predicted line profiles to get quantitative measures of the line
asymmetry which can be compared with the same measures for the observations.
We will then discuss the effects of outflows on the spectra and present some
detailed modeling of one source, L1527.
Based on these considerations, we will construct a truth table (Table 6) to
help decide which sources are good collapse candidates (\S 5).

\subsection {Evolutionary Modeling}

Zhou (1992) has modeled the behavior of CS lines as a function of time in
a collapsing cloud. He found that the CS lines get broader, weaker, and more
asymmetric as the collapse proceeds, as a result of the rising velocities
and dropping densities inside the infall radius.
Because HCO$^+$ has different optical depth and excitation
properties, the detailed behavior of its line profiles may differ from those
of CS. Consequently, we have modeled the evolution of the HCO$^{+}$ $J=3-2$
and $J=4-3$ line profiles,
as well as the H$^{13}$CO$^{+}$ $J=3-2$ and $J=4-3$ lines,
with time. We used the density and velocity fields of the inside-out collapse
model (Shu 1977) and a temperature distribution derived from scaling
that used for B335 (Zhou et al.\ 1990) to a
luminosity of 6.5 L$_{\sun}$, the average luminosity of the sources with
well-determined bolometric temperatures.
The sound speed, $a$, was set at 0.21 \kms.  Altering the sound speed would
slow down or speed up the evolution of the line profile.
Models were run for infall radii, $r_{inf}$,
of 0.005, 0.01, 0.02, 0.03, 0.04, 0.05, 0.06, and 0.08 pc, corresponding to
infall
times of 2.3$\times$10$^{4}$ yr for the earliest model and
3.8$\times$10$^{5}$ yr for the last.  The models had 30 shells, 15 inside
the infall radius and 15 outside the infall radius.
The outer radius, $r_{out}$,
was 0.2 pc.  Changing the outer radius affects the line profile very little.
We used a constant abundance of X(HCO$^{+}$) = 6$\times$10$^{-9}$.
The terrestrial abundance ratio of
X(HCO$^{+}$)/X(H$^{13}$CO$^{+}$) = 90 was assumed.   The
H$_2$-HCO$^{+}$ collisional rate coefficients are taken from Monteiro (1985)
for levels up to $J=4$; the rate coefficients for $J=5$ and 6 are scaled
appropriately from those calculated by Green (1975) for N$_{2}$H$^{+}$.

These parameters were used as input into a code which simulates the
Shu infall model and includes the temperature and abundance distribution.  The
resulting output, the density, velocity, kinetic temperature, turbulent width,
and X(HCO$^{+}$) at each shell, was then fed into the Monte Carlo
code described by Choi et al.\ (1995).
The output of this code, populations of each level of HCO$^+$ in each
shell, then became the input to a virtual telescope program that
simulated observations with the CSO, convolving the emission from each shell
with the spatial and velocity resolution of the actual observations (Figure 8).
 The
distance to the model cloud was 140 pc, which is the distance to the Taurus
Molecular Cloud and only
slightly farther than sources in the Ophiuchus cloud core ($d$ = 125$\pm$25 pc)
(de Geus et al.\ 1989).

The line profiles which emerge from these calculations are qualitatively
similar to those
of CS calculated by Zhou (1992), but some important differences emerge because
of the difference between CS and HCO$^{+}$.
The asymmetry between blue and red peaks increases with age, and it is stronger
than in the CS profiles. Unlike the CS models, the intensity of the HCO$^+$
$J=3-2$ line grows with time until $t \sim 1.9 \times 10^5$ yr
because the dropping opacity outweighs the
effects of dropping density. The $J=4-3$ line also grows initially, but
it decreases after $t = 9.4 \times 10^4$ yr since it is more sensitive to
density.
The H$^{13}$CO$^{+}$ $J=3-2$ and $J=4-3$ lines decrease with age and become
clearly weaker than the observations after a very short time and before the
blue-red asymmetry matches the observations. To get stronger H$^{13}$CO$^+$
lines while maintaining an isotope ratio of 90,
we increased the abundance to X(HCO$^{+}$) = 6$\times$10$^{-8}$ inside the
infall radius, while retaining X(HCO$^{+}$) = 6$\times$10$^{-9}$ in the
static envelope, based on some models of Rawlings et al.\ (1992), which  have
lower HCO$^{+}$ abundances in the outer envelope than in the interior.
These models (Figure 9)
can reproduce the observed H$^{13}$CO$^+$ lines, but the HCO$^+$
lines are now very opaque. One consequence is that the intensities of both
$J=3-2$ and $J=4-3$ lines increase with time, as the opacity drops, until
even later times.

Models were also run for a range of distances from 70 pc to 700 pc to check the
effect of distance on the observed line profiles, as observed with a beam of
fixed angular extent. The blue/red ratio decreases with distance from as high
as 3.52 to as low as 1.59 for our constant abundance model. At 700 pc, the line
profiles become noticeably asymmetric
only at the oldest times. The linewidth and line strength also decrease with
distance. All of these effects will make collapse harder to detect in more
distant clouds of low luminosity.
However, the more distant sources are often more luminous, and the
temperature distribution we used (appropriate for the average luminosity of
our sample)
will underestimate the line strength.  Because of the
dependence on distance, temperature, and abundance, we consider the models
presented here only as rough sketches of the evolution of a collapsing cloud.

\subsection {Quantitative Collapse Indicators}

We can use the models of line profiles to predict the behavior of various
quantities with time. For example, the asymmetry of the line profile can
be characterized in various ways. One is the ratio of the blue peak intensity
and the red peak intensity. The blue/red ratio is plotted versus time in
Figure 10 for both lines and both models of the abundance.
The blue/red ratio generally increases with time for both models of
abundance and both lines, but the values are clearly larger for the
model with constant abundance. Thus absolute values of the blue/red ratio
cannot be used to assign an age without further information on the
abundance.

For lines which are not clearly double-peaked, the blue/red ratio will
be undefined, but we can still characterize the asymmetry of the line
by a measure like the skewness.
The skewness is a non-dimensional quantity
defined as the ratio of the third moment of a distribution and the three-halves
power of the second moment, both normalized by the first moment.
We calculated the skewness as follows:
$$ Skewness = {
{\Sigma(T_A^{*}(v-v_{LSR})^{3}\delta v)\over{\Sigma(T_A^{*}\delta
v)}}\over{\Bigl({{\Sigma(T_A^{*}(v-v_{LSR})^{2}\delta
v)\over{\Sigma(T_A^{*}\delta v)}}}\Bigr)^{3/2}} },	\eqno (1)
$$
where $v_{LSR}$ is the central velocity of the source and $\delta v$ is the
channel width.
Line profiles with a negative skewness have a blue asymmetry,
and sources with a positive skewness have a red asymmetry.
The skewness of the model line profiles is also plotted in Figure 10.
The skewness rapidly becomes more negative with time, but eventually reaches
a maximum negative value.

We have measured the blue/red ratio and skewness of the observed HCO$^{+}$
$J=3-2$ and $J=4-3$ lines to see how these quantities compare to the
model values (see Table 5). Lines with a single peak, even if skewed, or three
peaks were assigned a
blue/red ratio of one.
Skewness was measured in the interval of $v_{LSR}$ $\pm$ 1.25 \kms.  The
$v_{LSR}$ of the source was calculated from the velocity centroid of the
optically thin
H$^{13}$CO$^{+}$ line.  This velocity interval was chosen because it
covered the central line core for all the sources while excluding as much
of the outflow wings as possible.  (Changing the velocity interval over which
the skewness is measured to 0.95 \kms\ or 2 \kms\ causes, at most, a 25\%
change in the skewness.  The sign is unchanged.)  As can be seen from Table 5
and Figure
11, all of the double-peaked blue asymmetric sources,
save the HCO$^{+}$ $J=4-3$ line in Serpens SMM2, have a negative skewness.
One of the two red asymmetric double-peaked sources, L483, has a
positive skewness in $J=4-3$ and zero skewness in $J=3-2$; and the
other, L1448-C, has a negative skewness.
It is possible to have a line with red asymmetry and negative or zero skewness
if the line profile displays strong blue wings.

In Figure 11, we plot the
skewness versus the blue/red ratio. The observed sources are represented by
pentagons and the models as triangles ($J=3-2$ line) or boxes
($J=4-3$ line) connected by solid or dashed lines, according to the
same scheme as in Fig. 10. The good collapse candidates (\S 4.3) are shown as
filled pentagons.  The sources with a blue/red ratio greater than
unity have negative skewness, as expected. The $J=4-3$ lines lie nicely
in the region of the models, but many of the $J=3-2$ lines have larger blue/red
ratios than are achieved in the models. In this plot, the difference
between the two models is small, except that the model with constant abundance
extends to much higher blue/red ratios and covers the region with quite
a few sources in the $J=4-3$ line. In Fig. 12, we show the blue/red ratio
versus the peak
line temperature. This plot reveals another difference between the data
and the models: a number of sources have considerably stronger lines than
are seen in either model. The arrow at the right indicates IRAS 16293,
which lies off the plot. Some of these sources have higher luminosity
and consequently higher temperatures than we have modeled; higher temperatures
would produce stronger lines. For the
$J=4-3$ line, we can imagine that intermediate values of abundance may
match the sources with blue/red ratios above two and line strengths
greater than 2 K, but the high blue/red ratios in the $J=3-2$ line
may be harder to explain. This figure also reveals that high peak
temperatures and high blue/red ratios are hard to achieve with the
models we have run probably because the assumed luminosity is lower than that
of many of the sources.

\subsection{Outflow versus Infall }

Many of the spectra in Figures 1--6 show substantial high velocity wings, which
are
not produced in the infall models of the last section.  Models with faster
infall or enhanced abundances of HCO$^+$ in the innermost parts of the infall
might produce stronger line wings, but these would be concentrated toward the
center of the infall. Maps of the distribution in space of the emission in
different velocity ranges can distinguish between infall and bipolar outflow.
The top two panels of Figure 13 are integrated intensity maps of one of our
infall models.  The simulated map, like the actual observations, was made with
15\arcsec\ spacing.  The integrated intensity of the two peaks are centered on
the infrared source and the red peak is more compact than the blue peak
(Fig. 13). The emission in the line wings should be very compact.  (The dashed
contours are the integrated intensity of the red-shifted peak and the red line
wing.)
Figure 13 also shows maps of the intensity integrated over different velocity
intervals in L1527.
The integrated blue and red line wing intensities are peaked on either side of
the source, indicating
that the line wings are due to outflow lobes and are clearly not related to
collapse.
The map of the velocity interval corresponding to the blue peak is fairly
circular
and peaks near the center position, but the red peak is somewhat displaced,
indicating some contamination by the outflow. Several other sources (notably
NGC 1333 IRAS 4A, Serpens SMM4, L483, and L1448-C) have bipolar outflow
emission in HCO$^+$, based on the maps of the line wings (Figures 13 to 15).

If bipolar outflows dominate the line wings, they might also dominate
the peaks. In this case a high blue/red ratio would not reflect collapse, but
rather the relative strength of the two bipolar lobes in the beam centered
on the source. In that case, we might expect an equal number of sources to
have stronger red peaks. Considering only the sources with two clearly
distinct peaks, there are 7 with stronger blue peaks and 2 with stronger
red peaks; for a sample of nine objects with intrinsically equal probability
for either peak to be stronger, the probability is only 0.07 that 7 will have
stronger blue peaks. Thus it is unlikely, but not impossible, that the
peaks are all caused by outflows.

Despite the statistical evidence that not all the double-peaked spectra are
caused by outflows, the chance of contamination by outflows is serious enough
that
we do not consider the HCO$^+$ data alone to be sufficient evidence for
collapse.
Instead we take a blue asymmetry in both HCO$^+$ $J=4-3$ and $J=3-2$ as an
indication of possible collapse worthy
of further investigation. Of the nine sources with blue asymmetry, three (B335,
IRAS 16293, and L1527) have been previously suggested as collapse candidates,
based on other lines.  Of the other six, Serpens SMM2 can probably be
eliminated from
consideration because the H$^{13}$CO$^{+}$ $J=3-2$ line has two velocity
components,
which have nearly the same velocities as the two peaks of the HCO$^{+}$ $J=3-2$
line.
For the others, we can compare the integrated intensity maps of the peaks
to those produced by an infall model.  In NGC 1333 IRAS 4A and 4B, the blue
peak is centered on IRAS 4A,
but the red peak increases in intensity to the north.
In Serpens SMM4, as in L1527, the two peaks are separated by about
10\arcsec\ (Fig. 14), in the same direction that the peaks are displaced in the
maps
of the line wings, suggesting contamination from outflow emission.
In HH25MMS and IRAS 20050, the maximum of one peak
stretches over a larger area than the other peak (Figure 15).  Since
rotation will also cause a shift of peaks, these sources cannot be ruled out
as collapse candidates, but more evidence would be needed.

It is noteworthy that both L483 and L1448-C,
which have blue/red ratios less than one, have strong bipolarity in the
integrated line wing emission (Fig. 15).
In L1448-C, the blue peak is displaced from the center by about 15\arcsec,
supporting our earlier suggestion that it is a feature in the outflow.
In L483, the red peak and the blue peak are displaced about 20\arcsec\ in an
east-west
direction, as are the peaks in the maps of the line wing emission.  Also, in
L483, the peaks are separated in the direction of the outflow.

Table 6 is a truth table, which includes information on whether the lines
have the correct asymmetry for collapse and whether the H$^{13}$CO$^+$ line
peaks in the self-absorption dip.
Table 6 also includes information on whether the integrated intensity maps of
the line peaks are compatible
with collapse, with a ``n" indicating that the maps of the line peaks are
displaced
in a way that suggest they are dominated by outflow.

\subsection{Detailed Modeling of L1527}

Given the uncertainties described above, we did not make detailed models of all
the sources with blue asymmetries.
Instead we focused on L1527, a source with relatively
good evidence for collapse (Zhou et al.\ 1994, Myers et al.\ 1995, Zhou et al.\
1996),
to see if the HCO$^+$ was consistent with previous models.
Using the procedure described in \S 4.1, we have modeled the spectra toward
L1527.
All models had an outer radius of 0.2 pc, a temperature distribution
appropriate
for the source luminosity (see \S 4.1), and constant abundance.
By comparing the model profiles to the observed H$^{13}$CO$^{+}$ $J=4-3$ and
$J=3-2$
line profiles, we constrained the abundance for a range of infall radii.
After scaling down the abundance in the H$^{13}$CO$^{+}$
models by a factor of 5, the predicted profiles for the
HC$^{18}$O$^{+}$ $J=4-3$ line are consistent with our upper limit of 0.07 K.
Scaling the abundances that fit the H$^{13}$CO$^+$ profiles up by
factors of 50 or 90 to model HCO$^{+}$, we found the model which best
matches the observed line profiles, focusing on the inner part of the
profiles to avoid the outflow (Figure 16).
This model has a constant abundance of X(HCO$^{+}$) =
2.5$\times$10$^{-8}$, 50 times the abundance of H$^{13}$CO$^+$. Models using
90 for the isotope ratio typically produced worse fits.  The absorption dip is
deeper in the model than in the observations, suggesting that a lower abundance
in the static envelope would fit better. We did not pursue this point, to avoid
introducing too many parameters. The best-fit model has an infall radius of
0.026 pc,
in agreement with modeling of single-dish H$_2$CO spectra and interferometer
data on $^{13}$CO by Zhou et al.\ (1996),
but larger than that found by Myers et al.\ (1995), r$_{inf}$ = 0.015 pc, using
different H$_2$CO data.
The wings of the observed lines are clearly not produced by an infall model,
and we attribute those to outflow as discussed earlier.  Also, we have noted
earlier that the integrated intensity maps of the line peaks show a
$\sim$10\arcsec\ displacement between the two peaks in an east-west direction,
as do the peaks in the maps of the line wing emission.  Although the peak
regions of the L1527 spectrum were used to determine the best-fitting model,
there is still much excess emission in the peak regions.  This emission is
probably contamination from the outflow and contributes to the displacement of
the two peaks seen in Figure 13.  Note that the large
blue/red ratio in the observations is not well-matched by the models, as
was also noted by Myers et al. (1995) in their modeling of the H$_{2}$CO
$J=2_{12}-1_{11}$ line.

\section {CONCLUSIONS}

We have surveyed 23 Class 0 sources in HCO$^{+}$ $J=3-2$ and $J=4-3$.
Nine sources (39\% of the sample) have a blue asymmetry, making them candidates
for protostellar collapse.  Three sources have a red asymmetry, and one source
has
different asymmetries in the two lines.  The evolutionary
progress of a collapsing core was modeled to provide predictions of what line
profiles and statistical quantities that characterize them would look like.
In general, the model line profiles are similar to the observed line profiles,
but
they have difficulty reproducing the large blue/red ratios seen in some
sources.

HCO$^{+}$ shows the collapse signature quite well in sources previously
identified as collapse candidates. According to chemical models, it may be
relatively abundant in collapsing envelopes where other molecules may deplete
(Rawlings et al.\ 1992). However, this molecule has some definite
disadvantages.
First, it emits strongly in outflowing gas, confusing collapse signatures.
Second, it is a difficult molecule to model because its lines can be quite
optically thick.
We conclude that it is a good molecule for surveys of possible collapse
candidates,
because its lines are strong, but confirmation with other molecular lines is
needed
before collapse can be definitely claimed.

Referring to Table 6, we find that B335, IRAS 16293, L1527 (all previously
suggested), as well as HH25MMS, Serpens SMM4, and IRAS 20050, are good
candidates on all the criteria.
NGC 1333 IRAS 4A and 4B meet the criteria other than the outflow
maps. Serpens SMM2 fails the criterion that H$^{13}$CO$^+$ peaks in the dip.

We reiterate that these six sources are only collapse {\it candidates}.  Before
we can definitely say that protostellar collapse is happening in these sources,
further work needs to be done.  To strengthen the case for collapse,
multi-transition observations are needed in other optically thick molecules
like
CS and H$_{2}$CO.  Also, two-dimensional modeling should be done to distinguish
between the effects of collapse and rotation.
\acknowledgments

We would like to thank Yangsheng Wang, Daniel Jaffe, Wenbin Li, and Byron
Mattingly for their help with observations.  This work was supported by NSF
grant AST-9317567.

\clearpage
\begin{planotable}{lllll}
\tablewidth{5in}
\tablenum{1}
\tablecaption{List of Class 0 Sources}
\tablehead{\colhead{Name} & \colhead{R.A.} & \colhead{Dec.} &
\colhead{Offpos} & \colhead{Reference}  \\
& \colhead{(1950.0)} & \colhead{(1950.0)} & \colhead{(\arcsec)} & }
\startdata
L1448-N & 03:22:31.5 & 30:34:49.0 & (--1200,0) & 1 \nl
L1448-C & 03:22:34.4 & 30:33:35.0 & (--1200,0) & 1 \nl
NGC 1333 IRAS 4A & 03:26:04.78 & 31:03:13.6 & (150,120) & 2 \nl
NGC 1333 IRAS 4B & 03:26:06.48 & 31:02:50.8 & (150,120) & 2 \nl
L1551-NE & 04:28:50.5 & 18:02:10 & (--600,0) & 3 \nl
L1527 & 04:36:49.3 & 25:57:16 & (--600,0) & 4 \nl
RNO43MM & 05:29:30.6 & 12:47:35.0 & (--900,0) & 5 \nl
NGC 2024 FIR 5 & 05:39:13.0 & --01:57:08.0 & (--1800,0) & 6 \nl
NGC 2024 FIR 6 & 05:39:13.7 & --01:57:30.0 & (--1800,0) & 6 \nl
HH25MMS & 05:43:34.0 & --00:14:42 & (--900,0) & 7 \nl
IRAS 08076 & 08:07:40.2 & --35:56:07.0 & (--900,0) & 8 \nl
VLA 1623 & 16:23:24.9 & --24:17:46.3 & (0,--900) & 9 \nl
IRAS 16293 & 16:29:20.9 & --24:22:13.0 & (0,--900) & 10 \nl
L483 & 18:14:50.6 & --04:40:49.0 & (--600,0) & 11 \nl
Serpens SMM1 & 18:27:17.3 & 01:13:23.0 & (--900,0) & 12 \nl
Serpens SMM4 & 18:27:24.7 & 01:11:10.0 & (--900,0) & 12 \nl
Serpens SMM3 & 18:27:27.3 & 01:11:55.0 & (--900,0) & 12 \nl
Serpens SMM2 & 18:27:28.0 & 01:10:45.0 & (--900,0) & 12 \nl
L723 & 19:15:42.0 & 19:06:49.0 & (--900,0) & 13 \nl
B335 & 19:34:35.4 & 07:27:24.0 & (--600,0) & 14 \nl
IRAS 20050 & 20:05:02.5 & 27:20:09.0 & (--900,0) & 15 \nl
S106FIR & 20:25:32.44 & 37:12:48.0 & (--900,0) & 16 \nl
L1157 & 20:38:39.6 & 67:51:33.0 & (--900,0) & 17
\tablecomments{References -- (1) Curiel et al.\ 1990 (2) Sandell et al.\ 1991
(3) Moriarty-Schieven et al.\ 1995a (4) Zhou et al.\ 1994
(5) Zinnecker et al.\ 1992  (6) Mezger et al.\ 1992 (7) Bontemps et al.\
1995 (8) Persi et al.\ 1990  (9) Andr\'e and
Montmerle 1994 (10) Walker et al.\ 1986
(11) Mardones et al.\ 1994 (12) Casali et al.\ 1993 (13) Davidson 1987 (14)
Chandler and Sargent 1993 (15) Bachiller et al.\ 1995 (16) Richer et al.\ 1993
(17) Tafalla and
Bachiller 1995}
\end{planotable}

\clearpage
\begin{planotable}{llllll}
\tablewidth{6.5in}
\tablenum{2}
\tablecaption{Bolometric Temperature of Class 0 Sources}
\tablehead{\colhead{Source} & \colhead{Range} & \colhead{T$_{bol}$} &
\colhead{Distance} & \colhead{L$_{bol}$} & \colhead{Reference}
\\ & \colhead{$\mu$m} & \colhead{K} & \colhead{pc} & \colhead{L$_{\sun}$} }
\startdata
L1448-C & 12-3500 & 55$\pm$9 & 300 & 6$\pm$0.4 & 1 \nl
NGC 1333 IRAS 4A & 50-2000 & 34$\pm$2 & 350 & 9.5$\pm$0.3 & 2,3 \nl
NGC 1333 IRAS 4B & 50-2000 & 36$\pm$2 & 350 & 8.4$\pm$0.9 & 2,3 \nl
L1551-NE & 12-1260 & 75$\pm$18 & 140 & 3.9$\pm$0.5 & 4,5 \nl
L1527 & 25-800 & 41$\pm$6 & 140 & 1.3$\pm$0.1 & 6,7 \nl
RNO43MM & 60-1300 & 36$\pm$4 & 400 & 4.5$\pm$0.3 & 6,8 \nl
IRAS 08076 & 12-1300 & 74$\pm$15 & 400 & 8.9$\pm$0.3 & 6,9 \nl
IRAS 16293 & 25-3000 & 42$\pm$0.7 & 125 & 11$\pm$0.1 & 6,10,11 \nl
L483 & 25-450 & 48$\pm$5 & 200 & 9.5$\pm$0.5 & 6,12 \nl
Serpens SMM1 & 20-2000 & 45$\pm$5 & 310 & 45$\pm$3 & 13,14,15\nl
Serpens SMM4 & 800-2000 & 35$\pm$5 & 310 & 3.9$\pm$0.3 & 13,14,15\nl
Serpens SMM3 & 60-1300 & 38$\pm$5 & 310 & 4.5$\pm$0.4 & 13,14,15 \nl
Serpens SMM2 & 60-1100 & 38$\pm$5 & 310 & 2.7$\pm$0.2 & 13,14,15 \nl
L723 & 25-1300 & 39$\pm$4 & 300 & 2.8$\pm$0.2 & 6,16,17 \nl
B335 & 60-1300 & 29$\pm$3 & 250 & 2.5$\pm$0.2 & 6,16,18,19,20,21 \nl
IRAS 20050 & 12-2700 & 69$\pm$7 & 700 & 206$\pm$13 & 6,22 \nl
\nl
VLA 1623 & 350-2000 & $\gtrsim$8.3$\pm$1.9 & 125 & $\gtrsim$0.06$\pm$0.009 &
11,23\nl
L1157 & 25-100 & $\lesssim$44$\pm$9 & 440 & $\gtrsim$6$\pm$0.6 & 6\nl
\nl
L1448-N & 12-100 & -- & 300 & -- & 6 \nl
NGC 2024 FIR 5 & 350-2700 & -- & 415 & -- & 24,25,26 \nl
NGC 2024 FIR 6 & 870-2700 & -- & 415 & -- & 24,25,26 \nl
HH25MMS & 52-1300 & -- & 400 & -- & 27,28 \nl
S106FIR & 450-1100 & -- & 600 & -- & 29 \nl

\tablecomments{References -- (1) Bachiller et al.\ 1991 (2) Sandell et al.\
1991
(3) Jennings et al. 1987 (4) Emerson et al.\ 1984
(5) Barsony and Chandler 1993 (6) IRAS Point Source Catalog
(7) Ladd et al.\ 1991  (8) Zinnecker et al.\ 1992 (9) Persi et al.\ 1994 (10)
Walker et al.\ 1990 (11) Andr\'e et al.\ 1990b (12) Fuller et al.\ 1995 (13)
McMullin et al.\ 1994 (14) Casali et al.\ 1993 (15) Hurt et al.\ 1996a (16)
Davidson 1987
(17) Cabrit and Andr\'e 1991 (18) Chandler et al.\ 1990 (19) Gee et al.\ 1985
(20) Keene et al.\ 1983 (21) Chandler and Sargent 1993
(22) Wilking et al.\ 1989b
(23) Andr\'e et al.\ 1993
(24) Mezger et al.\ 1988
(25) Mezger et al.\ 1992 (26) Wilson et al.\ 1995 (27) Cohen et al.\ 1984
(28) Bontemps et al.\ 1995 (29) Richer et al.\ 1993}
\end{planotable}

\clearpage
\begin{planotable}{llllll}
\tablewidth{6in}
\tablenum{3}
\tablecaption{List of Observed Lines}
\tablehead{\colhead{Molecule} & \colhead{Transition} & \colhead{Beamwidth} &
\colhead{$\eta_{mb}$} & \colhead{$\delta$v} & \colhead{Frequency} \\ & &
\colhead{(\arcsec)} & & \colhead{(\kms)} & \colhead{(GHz)}}
\startdata
H$^{18}$CO$^{+}$ & $J=3-2$ & 27 & 0.66 & 0.17 & 255.479389 \nl
H$^{13}$CO$^{+}$ & $J=3-2$ & 26 & 0.66 & 0.17 &  260.255478 \nl
HCO$^{+}$ & $J=3-2$ & 26 & 0.66 & 0.16 & 267.557620 \nl
H$^{13}$CO$^{+}$ & $J=4-3$ & 20 & 0.62 & 0.13 &  346.998540 \nl
HCO$^{+}$ & $J=4-3$ & 20 & 0.62 & 0.12 & 356.734288
\end{planotable}

\clearpage
\begin{planotable}{llllll}
\tablewidth{6in}
\tablenum{4}
\tablecaption{Results}
\tablehead{\colhead{Source} & \colhead{Molecule} & \colhead{Line} &
\colhead{$T_A^*$} & \colhead{Velocity} & \colhead{$\Delta$V} \\ & & &
\colhead{(K)} & \colhead{(\kms)} & \colhead{(\kms)}}
\startdata
L1448-N & HCO$^{+}$ & $J=3-2$ & 4.02$\pm$0.04 & 5.02$\pm$0.01 & 2.07$\pm$0.01
\nl
-- & H$^{13}$CO$^{+}$ & $J=3-2$ & 1.16$\pm$0.06 & 4.86$\pm$0.02 & 1.63$\pm$0.05
\nl
L1448-C & HCO$^{+}$ & $J=4-3$ & 1.26$\pm$0.10 & 4.38$\pm$0.07 & 2.02$\pm$0.13
\nl
-- & -- & -- & 2.69$\pm$0.10 & 5.59$\pm$0.07 & -- \nl
-- & -- & $J=3-2$ & 1.65$\pm$0.06 & 4.31$\pm$0.09 & 2.21$\pm$0.17  \nl
-- & -- & -- & 2.75$\pm$0.06 & 5.67$\pm$0.09 & --  \nl
-- & H$^{13}$CO$^{+}$ & $J=4-3$ & 0.69$\pm$0.06 & 5.45$\pm$0.02 &
0.78$\pm$0.05 \nl
-- & -- & $J=3-2$ & 0.99$\pm$0.02 & 5.31$\pm$0.01 &  0.83$\pm$0.02 \nl
NGC 1333 IRAS 4A & HCO$^{+}$ & $J=4-3$ & 4.23$\pm$0.07 & 6.61$\pm$0.07 &
3.36$\pm$0.13 \nl
-- & -- & -- & 1.03$\pm$0.07 & 8.22$\pm$0.07 & -- \nl
-- & -- & $J=3-2$ & 5.15$\pm$0.03 & 6.47$\pm$0.09 & 3.91$\pm$0.17 \nl
-- & -- & -- & 1.24$\pm$0.03 & 8.17$\pm$0.09 & -- \nl
-- & H$^{13}$CO$^{+}$ & $J=4-3$ & 0.49$\pm$0.05 & 6.99$\pm$0.03 &
1.25$\pm$0.08 \nl
-- & -- & $J=3-2$ & 0.73$\pm$0.05 & 7.06$\pm$0.02 &  1.21$\pm$0.06\nl
NGC 1333 IRAS 4B & HCO$^{+}$ & $J=4-3$ & 3.08$\pm$0.07 & 6.61$\pm$0.07 &
3.23$\pm$0.13 \nl
-- & -- & -- & 0.91$\pm$0.07 & 7.96$\pm$0.07 & -- \nl
-- & -- & $J=3-2$ & 3.73$\pm$0.09 & 6.47$\pm$0.09 & 3.40$\pm$0.17  \nl
-- & -- & -- & 0.88$\pm$0.09 & 8.00$\pm$0.09 & -- \nl
-- & H$^{13}$CO$^{+}$ & $J=4-3$ & $<0.08$ & -- & -- \nl
-- & & $J=3-2$ & 0.33$\pm$0.04 & 7.14$\pm$0.05 & 1.21$\pm$0.13 \nl
L1551-NE & HCO$^{+}$ & $J=3-2$ & 1.95$\pm$0.05 & 6.27$\pm$0.08 & 2.73$\pm$0.16
\nl
-- & -- & -- & 0.93$\pm$0.05 & 7.24$\pm$0.08 & -- \nl
-- & -- & -- & 0.82$\pm$0.05 & 8.04$\pm$0.08 & -- \nl
-- & H$^{13}$CO$^{+}$ & $J=3-2$ & 0.21$\pm$0.03 & 7.23$\pm$0.08 & 1.60$\pm$0.16
\nl
L1527 & HCO$^{+}$ & $J=4-3$ & 4.33$\pm$0.04 & 5.70$\pm$0.07 & 2.16$\pm$0.13 \nl
-- & -- & -- & 1.81$\pm$0.04 & 6.50$\pm$0.07 & -- \nl
-- & -- & $J=3-2$ & 5.00$\pm$0.04 & 5.51$\pm$0.09 & 2.38$\pm$0.17 \nl
-- & -- & -- & 2.47$\pm$0.04 & 6.36$\pm$0.09 & -- \nl
-- & H$^{13}$CO$^{+}$ & $J=4-3$ & 0.35$\pm$0.03 & 6.01$\pm$0.02 & 0.69$\pm$0.06
\nl
-- & & $J=3-2$ & 0.75$\pm$0.03 & 5.94$\pm$0.01 & 0.93$\pm$0.03 \nl
-- & H$^{18}$CO$^{+}$ & $J=4-3$ & $<0.07$ & -- & -- \nl
RNO43MM & HCO$^{+}$ & $J=3-2$ & 2.54$\pm$0.09 & 10.37$\pm$0.01 & 1.08$\pm$0.16
\nl
NGC 2024 FIR 5 & HCO$^{+}$ & $J=3-2$ & 8.22$\pm$0.08 & 11.74$\pm$0.01 &
1.70$\pm$0.16 \nl
-- & H$^{13}$CO$^{+}$ & $J=3-2$ & 2.08$\pm$0.07 & 11.38$\pm$0.01 &
1.84$\pm$0.04 \nl
NGC 2024 FIR 6 & HCO$^{+}$ & $J=3-2$ & 7.78$\pm$0.09 & 11.73$\pm$0.01 &
2.32$\pm$0.16 \nl
-- & H$^{13}$CO$^{+}$ & $J=3-2$ & 1.58$\pm$0.07 & 11.25$\pm$0.02 &
2.18$\pm$0.05 \nl
HH25MMS & HCO$^{+}$ & $J=4-3$ & 2.02$\pm$0.03 & 9.83$\pm$0.07 & 2.13$\pm$0.13
\nl
-- & -- & -- & 1.56$\pm$0.03 & 10.90$\pm$0.07 & -- \nl
-- & & $J=3-2$ & 3.53$\pm$0.05 & 10.02$\pm$0.01 & 2.21$\pm$0.01 \nl
-- & H$^{13}$CO$^{+}$ & $J=3-2$ & 0.84$\pm$0.05 & 10.34$\pm$0.02 &
1.19$\pm$0.05 \nl
IRAS 08076 & HCO$^{+}$ & $J=4-3$ & 1.95$\pm$0.06 & 6.28$\pm$0.01 &
1.60$\pm$0.13 \nl
-- & -- & $J=3-2$ & 4.53$\pm$0.12 & 6.19$\pm$0.01 & 1.19$\pm$0.17 \nl
-- & H$^{13}$CO$^{+}$ & $J=3-2$ & 1.22$\pm$0.06 & 6.48$\pm$0.01 & 0.91$\pm$0.03
\nl
VLA 1623 & HCO$^{+}$ & $J=4-3$ & 2.01$\pm$0.1 & 2.31$\pm$0.06 & 3.26$\pm$0.12
\nl
-- & -- & -- & 4.39$\pm$0.1 & 3.40$\pm$0.06 & -- \nl
-- & -- & -- & 4.62$\pm$0.1 & 4.24$\pm$0.06 & -- \nl
-- & -- & $J=3-2$ & 3.42$\pm$0.06 & 2.19$\pm$0.08 & 2.89$\pm$0.16 \nl
-- & -- & -- & 3.96$\pm$0.06 & 3.32$\pm$0.08 & -- \nl
-- & -- & -- & 4.36$\pm$0.06 & 4.28$\pm$0.08 & -- \nl
-- & H$^{13}$CO$^{+}$ & $J=3-2$ & 1.10$\pm$0.05 & 3.98$\pm$0.01 & 1.03$\pm$0.03
\nl
IRAS 16293 & HCO$^{+}$ & $J=4-3$ & 15.09$\pm$0.14 & 3.48$\pm$0.06 &
4.10$\pm$0.12 \nl
-- & -- & -- & 4.76$\pm$0.14 & 4.86$\pm$0.06 & -- \nl
-- & -- & $J=3-2$ & 17.60$\pm$0.07 & 3.31$\pm$0.08 & 3.37$\pm$0.16 \nl
-- & -- & -- & 6.66$\pm$0.07 & 4.92$\pm$0.08 & -- \nl
-- & H$^{13}$CO$^{+}$ & $J=3-2$ & 1.52$\pm$0.07 & 3.65$\pm$0.17 & 2.48$\pm$0.34
\nl
-- & -- & -- & 1.33$\pm$0.07 & 4.81$\pm$0.17 & -- \nl
-- & HC$^{18}$O$^{+}$ & $J=3-2$ & 0.18$\pm$0.05 & 4.52$\pm$0.08 &
2.7$\pm$0.31\nl
L483 & HCO$^{+}$ & $J=4-3$ & 1.12$\pm$0.13 & 4.87$\pm$0.06 & 2.65$\pm$0.12 \nl
-- & -- & -- & 3.48$\pm$0.13 & 5.71$\pm$0.06 & -- \nl
-- & -- & $J=3-2$ & 1.47$\pm$0.04 & 4.77$\pm$0.08 & 2.33$\pm$0.16 \nl
-- & -- & -- & 3.10$\pm$0.04 & 5.86$\pm$0.08 & -- \nl
-- & H$^{13}$CO$^{+}$ & $J=3-2$ & 0.97$\pm$0.04 & 5.57$\pm$0.01 & 0.90$\pm$0.03
\nl
-- & HC$^{18}$O$^{+}$ & $J=3-2$ & 0.11$\pm$0.01 & 5.31$\pm$0.06 & 1.08$\pm$0.23
\nl
Serpens SMM1 & HCO$^{+}$ & $J=4-3$ & 2.14$\pm$0.14 & 7.13$\pm$0.06 &
3.87$\pm$0.12 \nl
-- & -- & -- & 2.17$\pm$0.14 & 9.79$\pm$0.06 & -- \nl
-- & -- & $J=3-2$ & 2.30$\pm$0.12 & 7.26$\pm$0.08 & 4.19$\pm$0.15 \nl
-- & -- & -- & 2.40$\pm$0.12 & 9.74$\pm$0.08 & -- \nl
-- & H$^{13}$CO$^{+}$ & $J=3-2$ & 1.19$\pm$0.06 & 8.53$\pm$0.02 & 1.55$\pm$0.04
\nl
-- & HC$^{18}$O$^{+}$ & $J=3-2$ & 0.24$\pm$0.02 & 8.44$\pm$0.04 & 1.47$\pm$0.11
\nl
Serpens SMM4 & HCO$^{+}$ & $J=4-3$ & 4.56$\pm$0.08 & 7.66$\pm$0.06 &
3.98$\pm$0.12 \nl
-- & -- & -- & 1.73$\pm$0.08 & 9.10$\pm$0.06 & 3.98$\pm$0.12 \nl
-- & -- & $J=3-2$ & 6.58$\pm$0.04 & 7.24$\pm$0.08 & 6.58$\pm$0.16 \nl
-- & -- & -- & 2.16$\pm$0.04 & 9.41$\pm$0.08 & -- \nl
-- & H$^{13}$CO$^{+}$ & $J=3-2$ & 0.95$\pm$0.04 & 8.02$\pm$0.02 & 1.56$\pm$0.05
\nl
-- & HC$^{18}$O$^{+}$ & $J=3-2$ & 0.15$\pm$0.02 & 7.65$\pm$0.04 & 0.92$\pm$0.1
\nl
Serpens SMM3 & HCO$^{+}$ & $J=4-3$ & 3.02$\pm$0.16 & 7.83$\pm$0.02 &
1.45$\pm$0.12 \nl
-- & -- & $J=3-2$ & 5.00$\pm$0.06 & 7.74$\pm$0.01 & 1.46$\pm$0.11 \nl
Serpens SMM2 & HCO$^{+}$ & $J=4-3$ & 1.68$\pm$0.08 & 6.93$\pm$0.06 &
2.53$\pm$0.12 \nl
-- & -- & -- & 1.06$\pm$0.08 & 8.02$\pm$0.06 & -- \nl
-- & -- & $J=3-2$ & 2.75$\pm$0.05 & 6.66$\pm$0.06 & 2.81$\pm$0.11 \nl
-- & -- & -- & 2.37$\pm$0.05 & 7.89$\pm$0.06 & -- \nl
-- & H$^{13}$CO$^{+}$ & $J=3-2$ & 0.55$\pm$0.03 & 7.08$\pm$0.08 & 1.82$\pm$0.16
\nl
-- & -- & -- & 0.32$\pm$0.03 & 8.23$\pm$0.08 & -- \nl
-- & HC$^{18}$O$^{+}$ & $J=3-2$ & 0.07$\pm$0.02 & 6.62$\pm$0.08 & 0.81$\pm$0.15
\nl
L723 & HCO$^{+}$ & $J=4-3$ & 1.09$\pm$0.15 & 11.02$\pm$0.05 & 1.56$\pm$0.12 \nl
-- & -- & $J=3-2$ & 1.79$\pm$0.05 & 11.06$\pm$0.01 & 1.93$\pm$0.16 \nl
B335 & HCO$^{+}$ & $J=4-3$ & 2.37$\pm$0.07 & 8.17$\pm$0.06 & 1.69$\pm$0.12 \nl
-- & -- & -- & 1.05$\pm$0.07 & 8.78$\pm$0.06 & -- \nl
-- & -- & $J=3-2$ & 3.31$\pm$0.06 & 8.00$\pm$0.08 & 1.39$\pm$0.16 \nl
-- & -- & -- & 1.59$\pm$0.06 & 8.78$\pm$0.08 & -- \nl
-- & H$^{13}$CO$^{+}$ & $J=3-2$ & 0.91$\pm$0.07 & 8.41$\pm$0.02 & 0.62$\pm$0.04
\nl
-- & HC$^{18}$O$^{+}$ & $J=3-2$ & 0.09$\pm$0.02 & 8.25$\pm$0.04 & 0.60$\pm$0.11
\nl
IRAS 20050 & HCO$^{+}$ & $J=4-3$ & 3.12$\pm$0.08 & 5.93$\pm$0.01 &
2.89$\pm$0.12 \nl
-- & -- & $J=3-2$ & 3.26$\pm$0.08 & 5.27$\pm$0.05 & 4.51$\pm$0.11 \nl
-- & -- & -- & 2.32$\pm$0.08 & 7.30$\pm$0.05 & -- \nl
-- & H$^{13}$CO$^{+}$ & $J=3-2$ & 0.42$\pm$0.05 & 6.09$\pm$0.06 & 2.56$\pm$0.14
\nl
S106FIR & HCO$^{+}$ & $J=4-3$ & 4.55$\pm$0.16 & -1.07$\pm$0.02 & 3.50$\pm$0.12
\nl
-- & -- & $J=3-2$ & 4.14$\pm$0.11 & -1.27$\pm$0.02 & 3.22$\pm$0.16 \nl
L1157 & HCO$^{+}$ & $J=4-3$ & 1.17$\pm$0.08 & 2.34$\pm$0.06 & 1.69$\pm$0.12 \nl
-- & -- & -- & 1.51$\pm$0.08 & 3.18$\pm$0.06 & -- \nl
-- & -- & $J=3-2$ & 2.03$\pm$0.08 & 2.25$\pm$0.08 & 1.93$\pm$0.16 \nl
-- & -- & -- & 1.78$\pm$0.08 & 3.22$\pm$0.08 & -- \nl
-- & H$^{13}$CO$^{+}$ & $J=3-2$ & 0.31$\pm$0.07 & 2.83$\pm$0.09 & 1.52$\pm$0.20
\end{planotable}

\clearpage
\begin{planotable}{llll}
\tablewidth{6in}
\tablenum{5}
\tablecaption{Various Collapse Indicators for HCO$^{+}$}
\tablehead{\colhead{Source} & \colhead{Line} &  \colhead{Skewness} &
\colhead{Blue/Red}}
\startdata
L1448-N & $J=3-2$ & 0.18$\pm$0.29 & 1.00 \nl
L1448-C & $J=4-3$ & -0.50$\pm$0.42 & 0.47 \nl
-- & $J=3-2$ & -0.26$\pm$0.42 & 0.60   \nl
NGC 1333 IRAS 4A & $J=4-3$ & -0.33$\pm$0.24 & 4.09  \nl
-- & $J=3-2$ & -0.65$\pm$0.27 & 4.17 \nl
NGC 1333 IRAS 4B & $J=4-3$ & -0.56$\pm$0.22 & 3.37  \nl
-- & $J=3-2$ & -0.45$\pm$0.25 & 4.22  \nl
L1551-NE & $J=3-2$ & -0.34$\pm$0.21 & 1.00 \nl
L1527 & $J=4-3$ & -0.54$\pm$0.19 & 2.39  \nl
-- & $J=3-2$ & -0.40$\pm$0.22 & 2.02  \nl
RNO43MM & $J=3-2$ & -0.58$\pm$0.53 & 1.00  \nl
NGC 2024 FIR 5 & $J=3-2$ & 0.42$\pm$0.20 & 1.00  \nl
NGC 2024 FIR 6 & $J=3-2$ & 0.66$\pm$0.20 & 1.00  \nl
HH25MMS & $J=4-3$ & -0.18$\pm$0.22 & 1.30  \nl
-- & $J=3-2$ & -0.44$\pm$0.20 & 1.00  \nl
IRAS 08076 &  $J=4-3$ & -0.46$\pm$0.25 & 1.00  \nl
-- & $J=3-2$ & -0.94$\pm$0.38 & 1.00  \nl
VLA 1623 & $J=4-3$ & -0.97$\pm$0.23 & 1.00 \nl
-- & $J=3-2$ & -0.86$\pm$0.27 & 1.00  \nl
IRAS 16293 & $J=4-3$ & -0.66$\pm$0.11 & 3.17  \nl
-- & $J=3-2$ & -0.38$\pm$0.14 & 2.64  \nl
L483 & $J=4-3$ & 0.15$\pm$0.3 & 0.32  \nl
-- & $J=3-2$ & -0.01$\pm$0.28 & 0.47  \nl
Serpens SMM1 & $J=4-3$ & -0.22$\pm$0.10 & 0.99  \nl
-- & $J=3-2$ & 0.09$\pm$0.12 & 0.96  \nl
Serpens SMM4 & $J=4-3$ & -0.41$\pm$0.15 & 2.64  \nl
-- & $J=3-2$ & -0.83$\pm$0.19 & 3.05  \nl
Serpens SMM3 & $J=4-3$ & -0.20$\pm$0.24 & 1.00  \nl
-- & $J=3-2$ & -0.48$\pm$0.16 & 1.00 \nl
Serpens SMM2 & $J=4-3$ & 0.08$\pm$0.16 & 1.63  \nl
-- & $J=3-2$ & -0.27$\pm$0.12 & 1.16  \nl
L723 & $J=4-3$ & -0.005$\pm$0.44 & 1.00  \nl
-- & $J=3-2$ & 0.05$\pm$0.30 & 1.00  \nl
B335 & $J=4-3$ & -0.15$\pm$0.46 & 2.25  \nl
-- & $J=3-2$ & -0.46$\pm$0.53 & 2.08  \nl
IRAS 20050 & $J=4-3$ & -0.15$\pm$0.13 & 1.00  \nl
-- & $J=3-2$ & -0.13$\pm$0.10 & 1.41  \nl
S106FIR & $J=4-3$ & -0.11$\pm$0.14 & 1.00  \nl
-- & $J=3-2$ & -0.15$\pm$0.24 & 1.00  \nl
L1157 & $J=4-3$ & -0.37$\pm$0.30 & 0.77  \nl
-- & $J=3-2$ & -0.25$\pm$0.23 & 1.14
\end{planotable}

\clearpage
\begin{planotable}{lllll}
\tablewidth{6.5in}
\tablenum{6}
\tablecaption{Truth Table}
\tablehead{\colhead{Source} & \colhead{Asymmetry 3-2} & \colhead{Asymmetry 4-3}
& \colhead{H$^{13}$CO$^{+}$ peaks in dip} & \colhead{Line peak maps}}
\startdata
L1448-N & r\tablenotemark{a} & --\tablenotemark{b} & y\tablenotemark{c} & --
\nl
L1448-C & r & r & close\tablenotemark{d} & n\tablenotemark{e} \nl
NGC 1333 IRAS 4A & b\tablenotemark{f} & b & close & n \nl
NGC 1333 IRAS 4B & b & b & y & n \nl
L1551-NE & b & -- & n & -- \nl
L1527 & b & b & y & y \nl
RNO43MM & n & -- & -- & -- \nl
NGC 2024 FIR 5 & n & -- & n & -- \nl
NGC 2024 FIR 6 & n & -- & n & -- \nl
HH25MMS & b & b & y & ?\tablenotemark{g} \nl
IRAS 08076 & n & n & n & -- \nl
VLA 1623 & r & r & y & ? \nl
IRAS 16293 & b & b & y & y \nl
L483 & r & r & y & n \nl
Serpens SMM1 & n & n & y & -- \nl
Serpens SMM4 & b & b & close & y \nl
Serpens SMM2 & b & b & n & ? \nl
Serpens SMM3 & n & n & -- & -- \nl
L723 & n & n & -- & -- \nl
B335 & b & b & y & -- \nl
IRAS 20050 & b & b & y & ? \nl
S106FIR & n & n & -- & -- \nl
L1157 & b & r & y & n  \nl
\tablenotetext{a}{r means red asymmetry.}
\tablenotetext{b}{- means no observations.}
\tablenotetext{c}{y means the source satisfies the criterion.}
\tablenotetext{d}{close means the H$^{13}$CO$^{+}$ peaks between a peak and the
dip of the HCO$^{+}$ line.}
\tablenotetext{e}{n means the source does not satisfy the criterion.}
\tablenotetext{f}{b means blue asymmetry.}
\tablenotetext{g}{? means the peaks overlap but one is spread over a larger
area than the other.}
\end{planotable}

\clearpage
\begin{figure}
\caption{HCO$^{+}$ and H$^{13}$CO$^{+}$ $J=3-2$ spectra toward the center
of nine Class 0 sources.  These sources all show a blue asymmetry in the
HCO$^{+}$ $J=3-2$ line characteristic of collapse.  The solid line is the
HCO$^{+}$ spectrum and the dashed line is the H$^{13}$CO$^{+}$ spectrum.
The scale for the spectra of IRAS 16293 is shown on the right side of its
panel. The H$^{13}$CO$^{+}$ $J=3-2$ spectrum for HH25MMS is from the
position (0,6).}
\end{figure}

\begin{figure}
\caption{HCO$^{+}$ and, where available,
H$^{13}$CO$^{+}$ $J=4-3$ spectra toward the center
of nine Class 0 sources.  These sources all show a blue asymmetry in the
HCO$^{+}$ $J=4-3$ line characteristic of collapse.  The solid line is the
HCO$^{+}$ spectrum and the dashed line is the H$^{13}$CO$^{+}$ spectrum.
The scale for the spectrum of IRAS 16293 is shown on the right side of its
panel.  }
\end{figure}

\begin{figure}
\caption{HCO$^{+}$ and H$^{13}$CO$^{+}$ $J=3-2$ spectra toward the center
of four Class 0 sources.  The first three sources, VLA 1623, L1448-C, and L483,
all show a red asymmetry in the
HCO$^{+}$ $J=3-2$.  L1157 has blue asymmetry in HCO$^{+}$ $J=3-2$.  The solid
line is the
HCO$^{+}$ spectrum and the dashed line is the H$^{13}$CO$^{+}$ spectrum.}
\end{figure}

\begin{figure}
\caption{HCO$^{+}$ and, where available, H$^{13}$CO$^{+}$ $J=4-3$ spectra
toward the center
of four Class 0 sources.  The first three sources, VLA 1623, L1448-C, and L483,
also have a red asymmetry in the
HCO$^{+}$ $J=3-2$.  L1157 which has blue asymmetry in HCO$^{+}$ $J=3-2$ shows
a red signature in $J=4-3$.  The solid line is the
HCO$^{+}$ spectrum and the dashed line is the H$^{13}$CO$^{+}$ spectrum.}
\end{figure}

\begin{figure}
\caption{HCO$^{+}$ and, where available, H$^{13}$CO$^{+}$ $J=3-2$ spectra
toward the center
of ten Class 0 sources.  Most of these sources do not show any clear asymmetry
in
the HCO$^{+}$ $J=3-2$ line.  The solid line is the
HCO$^{+}$ spectrum and the dashed line is the H$^{13}$CO$^{+}$ spectrum.}
\end{figure}

\begin{figure}
\caption{HCO$^{+}$ $J=4-3$ spectra toward the center
of five Class 0 sources.  These sources do not show any clear asymmetry in
the HCO$^{+}$ $J=4-3$ line.}
\end{figure}

\begin{figure}
\caption{H$^{13}$CO$^{+}$ and HC$^{18}$O$^{+}$ $J=3-2$ spectra toward the
center
of six Class 0 sources.  The solid line is the
H$^{13}$CO$^{+}$ spectrum and the dashed line is the H$^{18}$CO$^{+}$ spectrum.
In all of these sources, the HC$^{18}$O$^{+}$ $J=3-2$ has been scaled up by 2
except for L483, SMM2 and B335 where the HC$^{18}$O$^{+}$ $J=3-2$ has
been scaled up by a
factor of 4.}
\end{figure}

\clearpage
\begin{figure}
\caption{The top panel is a set of 8 models following the temporal evolution of
the HCO$^{+}$ (solid) and H$^{13}$CO$^{+}$ (dashed) $J=3-2$ line profiles for
a constant abundance X(HCO$^{+}$) = 6$\times$10$^{-9}$.  The
bottom panel shows the temporal evolution of the
HCO$^{+}$ and H$^{13}$CO$^{+}$ $J=4-3$ line profiles.
The times since infall began are given in the upper panels and are the
same in the lower panels.}
\end{figure}

\begin{figure}
\caption{The same as the previous figure, but with
 X(HCO$^{+}$) = 6$\times$10$^{-8}$ inside the infall radius
and X(HCO$^{+}$) = 6$\times$10$^{-9}$ in the static envelope outside the infall
radius.  }
\end{figure}

\begin{figure}
\caption{The top panel is the skewness of our evolutionary line profiles versus
time.  The bottom panel is the blue/red ratio of our evolutionary line
profiles versus time. A Gaussian line would have a skewness of 0 and a
blue/red ratio of 1 (the dash-dotted line).}
\end{figure}

\begin{figure}
\caption{The top panel is the skewness of our observed line profiles versus
the blue/red ratio for the HCO$^{+}$ $J=3-2$ line.  The bottom
panel is the skewness of our observed line profiles versus
the blue/red ratio for the HCO$^{+}$ $J=4-3$ line.  The triangular and the
square points are the $J=3-2$ and $4-3$ lines from the
models.  The spectra from Figure 8 are connected by a dashed line; those from
Figure 9 by a solid line.  The open and filled pentagonal points are the Class
0 sources and the good collapse candidates (see \S5), respectively.  A blue/red
value of 1 was was assigned to single-  or triple-peaked lines.}
\end{figure}

\begin{figure}
\caption{The top panel is the blue/red ratio of our observed line
profiles versus
the peak temperature for the HCO$^{+}$ $J=3-2$ line.  The bottom
panel is the blue/red ratio of our observed line profiles versus
the peak temperature for the HCO$^{+}$ $J=4-3$ line.  The triangular and the
square points are the $J=3-2$ and $4-3$ lines from the
models.  The spectra from Figure 8 are connected by a dashed line; those from
Figure 9 by a solid line.  The open and filled pentagonal points are the Class
0 sources and the good collapse candidates (see \S5), respectively.  A blue/red
value of 1 was was assigned to single- or triple-peaked lines.}
\end{figure}

\begin{figure}
\caption{The top two plots are integrated intensity maps of the red and blue
peaks and the blue and red lobes of the HCO$^{+}$ $J=3-2$ line in a simulation
of a collapsing cloud.  The velocity intervals are -0.43 to -0.27 \kms\ and
0.27
to 0.43 \kms\ for the peaks
and -3.09 to -0.43 \kms\ and 0.43 to 3.09 \kms\ for the lobes.  The contour
level spacing is 0.1 then 0.2 K \kms\ thereafter. In all plots,
the solid contours are the blueshifted peak and the blue line wing and the
dashed
contours are the red-shifted peak and the red line wing, respectively.
The middle two plots are integrated intensity maps of the red and blue peaks
and the blue and red lobes of the HCO$^{+}$ $J=3-2$ line in L1527.  The
velocity intervals are 5 to 5.7 \kms and 6.1 to 7 \kms for the peaks
and 2 to 5 \kms and 7 to 10 \kms for the lobes.  The contour level spacing is 1
K \kms\ for the peaks and 0.5 K \kms\ for the lobes.  The black square is the
location of the IRAS source.  The bottom two plots are integrated intensity
maps of the red and blue peaks and the blue and red lobes of the HCO$^{+}$
$J=3-2$ line in NGC 1333 IRAS 4A.  The velocity intervals are 4.3 to 5.09 \kms\
and 5.7 to 6.33 \kms \ for the peaks
and 2.6 to 4.3 \kms\ and 6.33 to 8.97 \kms\ for the lobes.  The contour level
spacing is: 0.5 K \kms\ for the blue peak and blue lobe, 0.25 then 0.5 K \kms\
thereafter for the red peak, and 0.1 K \kms\ for the red lobe.  The two black
squares are the location of the submillimeter sources detected by Sandell et
al.\ (1991).  The source at (0,0) is NGC 1333 IRAS 4A and the source at
(22,-23) is NGC 1333 IRAS 4B.}
\end{figure}

\begin{figure}
\caption{The top two plots are integrated intensity maps of the red and blue
peaks and the blue and red lobes of the HCO$^{+}$ $J=3-2$ line in IRAS
20050+2720.  The velocity intervals are 4.62 to 5.91 \kms\ and 6.65
to 7.95 \kms\ for the peaks
and -0.85 to 4.62 \kms\ and 7.95 to 10.74 \kms\ for the lobes.  The contour
level spacing is: 0.5 K \kms\ for the peaks, 1 then 0.5 K \kms\ thereafter for
the blue lobe, and 0.4 K \kms\ for the red lobe.  The black square is the
location of the IRAS source.  In all plots,
the solid contours are the blueshifted peak and the blue line wing and the
dashed
contours are the red-shifted peak and the red line wing, respectively.
The middle two plots are integrated intensity maps of the red and blue peaks
and the blue and red lobes of the HCO$^{+}$ $J=3-2$ line in SMM4.  The velocity
intervals are 6.3 to 8.02 \kms\ and 9.09 to 10.03 \kms\ for the peaks
and 2.42 to 6.3 \kms\ and 10.03 to 13.13 \kms\ for the lobes.  The contour
level spacing is: 1 K \kms\ for the blue peak, 0.25 K \kms\ for the red peak,
0.5 K \kms\ for the blue lobe, and 0.3 K \kms\ for the red lobe.  The black
square is the location of the submillimeter continuum source (Casali et al.\
1993).  The bottom two plots are integrated intensity maps of the red and blue
peaks and the blue and red lobes of the HCO$^{+}$ $J=4-3$ line in HH25MMS.  The
velocity intervals are 9.3 to 10.11 \kms\ and 10.51 to 11.3 \kms\ for the peaks
and 7.17 to 9.3 \kms\ and 11.3 to 13.03 \kms\ for the lobes.  The contour level
spacing is: 0.25 K \kms\ for the blue peak, 0.1 K \kms\ for the red peak, 0.18
K \kms\ for the blue lobe, and 0.15 K \kms\ for the red lobe.  The black square
is the location of the 1.3 mm radio continuum source (Bontemps et al.\ 1995).
}
\end{figure}

\begin{figure}
\caption{The top two plots are integrated intensity maps of the red and blue
peaks and the blue and red lobes of the HCO$^{+}$ $J=3-2$ line in L483.  The
velocity intervals are 4.3 to 5.09 \kms\ and 5.7
to 6.33 \kms\ for the peaks
and 2.6 to 4.3 \kms\ and 6.33 to 8.97 \kms\ for the lobes.  The contour level
spacing is: 0.2 K \kms\ for the blue peak, 0.25 K \kms\ for the red peak, 0.2
then 0.1 K \kms\ thereafter for the lobes.  The black square is the location of
the IRAS source.  In all plots,
the dashed
contours are the red-shifted peak and the red wing, respectively.
The bottom two plots are integrated intensity maps of the red and blue peaks
and the blue and red lobes of the HCO$^{+}$ $J=3-2$ line in L1448-C.  The
velocity intervals are 3.97 to 4.49 \kms\ and 5.16 to 6.18 \kms\ for the peaks
and 0.57 to 3.97 \kms\ and 6.18 to 7.55 \kms\ for the lobes.  The contour level
spacing is: 0.3 then 0.1 K \kms\ thereafter for the blue peak and red lobe, 0.4
K \kms\ for the red peak, 0.2 K \kms\ for the blue lobe.  The black square is
the location of the 2 cm radio continuum source (Curiel et al.\ 1990).}
\end{figure}

\begin{figure}
\caption{The top left panel is the observed HCO$^{+}$ $J=3-2$ line profile
(solid
line) and the simulated line profile (dashed line).  The infall model has
$r_{inf}$ = 0.026 pc and X(HCO$^{+}$) = 2.1$\times$10$^{-8}$.  The vertical
lines indicate the velocity intervals used in Figure 13.  The top right
panel is the observed HCO$^{+}$ $J=4-3$ line profile (solid
line) and the simulated line profile (dashed line) for the same model.  The
middle left panel is the observed H$^{13}$CO$^{+}$ $J=3-2$ line profile (solid
line) and the simulated line profile (dashed line).  The infall model has
X(H$^{13}$CO$^{+}$) = 4.1$\times$10$^{-10}$.  The middle right panel is the
observed H$^{13}$CO$^{+}$ $J=4-3$ line profile (solid
line) and the simulated line profile (dashed line).  The bottom left panel is
the observed HC$^{18}$O$^{+}$ $J=3-2$ line profile (solid
line) and the simulated line profile (dashed line).  The infall model has
X(HC$^{18}$O$^{+}$) = 7.4$\times$10$^{-11}$.  The bottom right panel is the
observed HC$^{18}$O$^{+}$ $J=4-3$ line profile (solid
line) and the simulated line profile (dashed line).}
\end{figure}

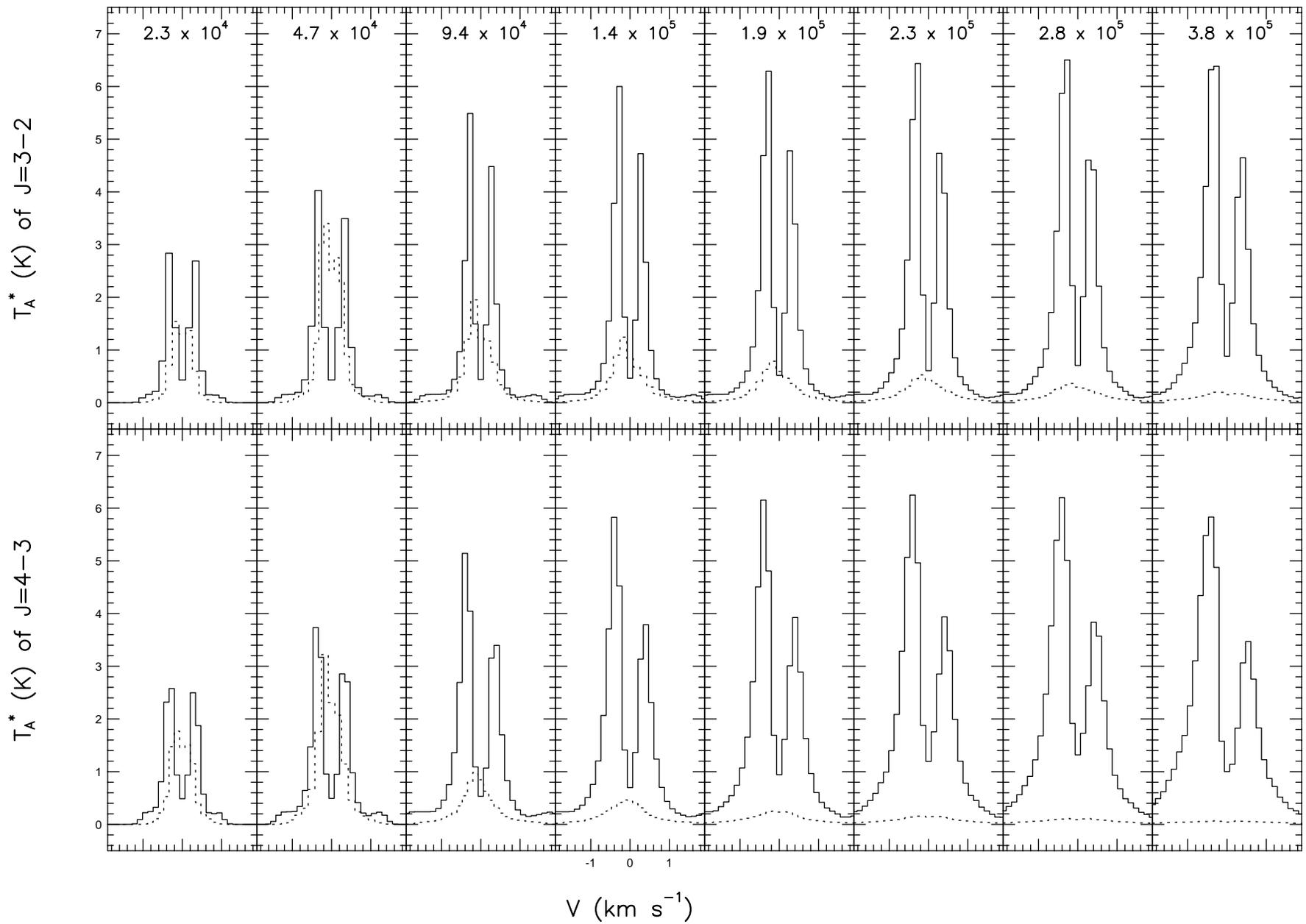
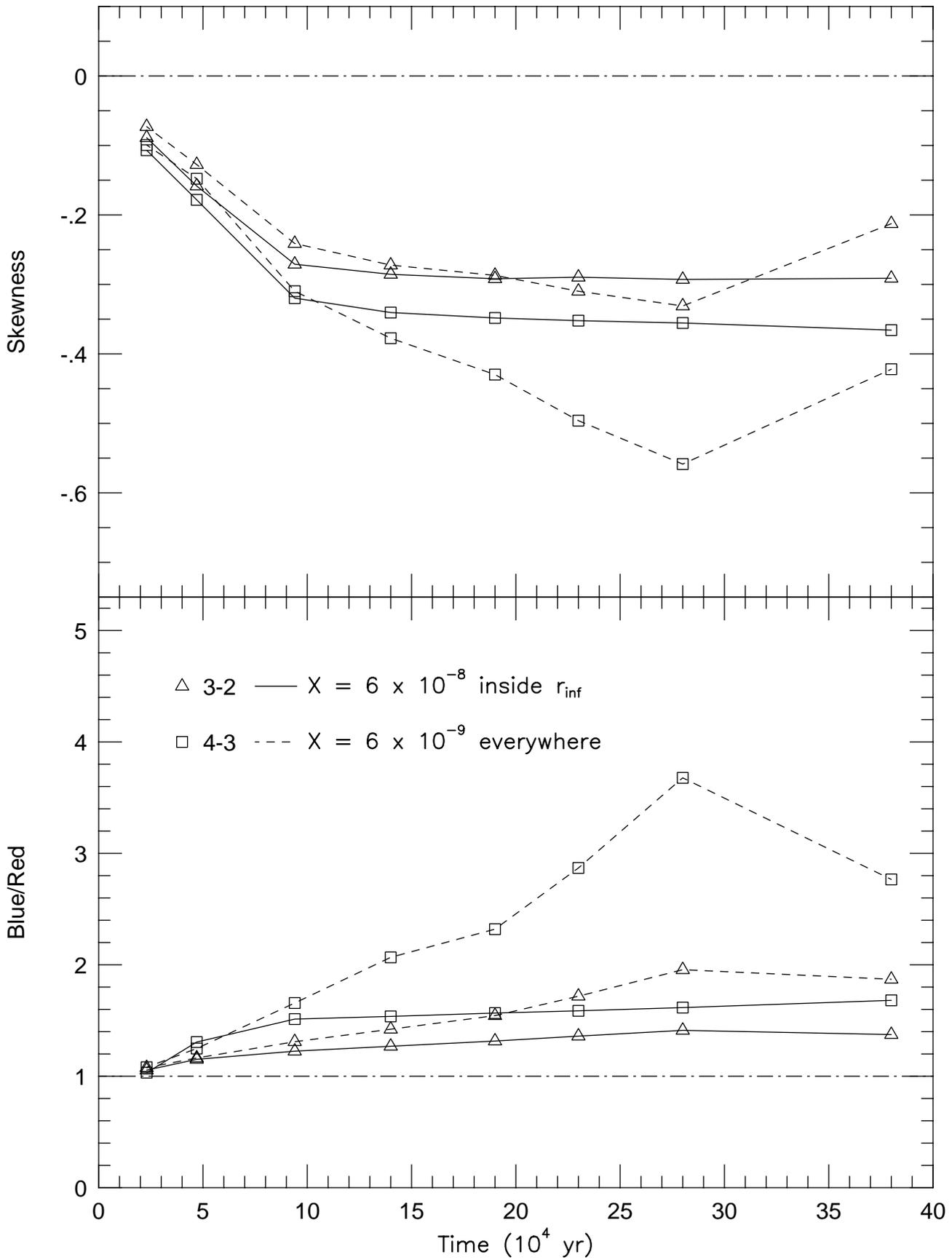
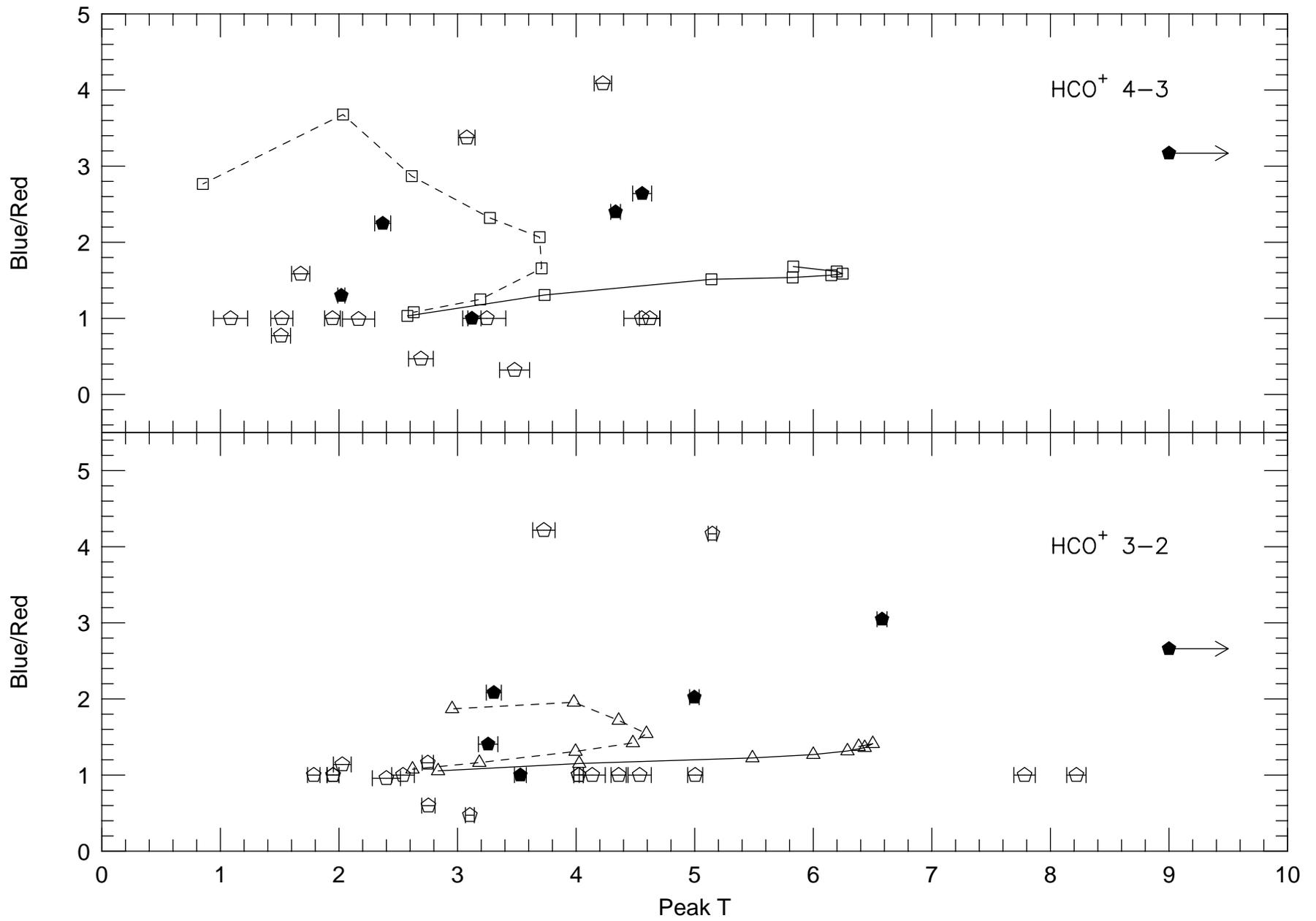
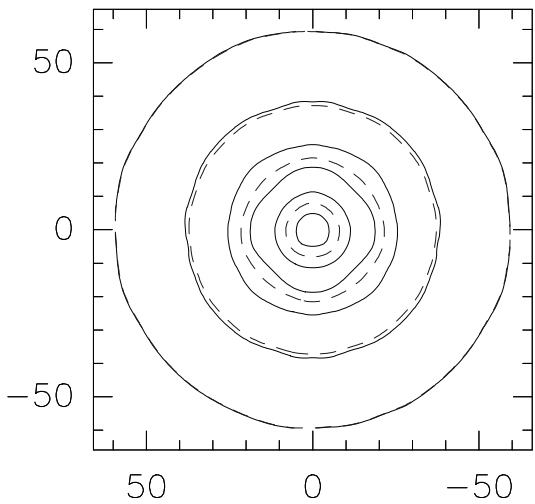
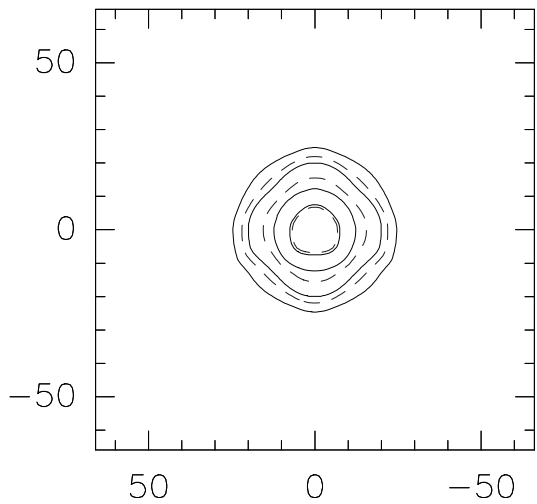
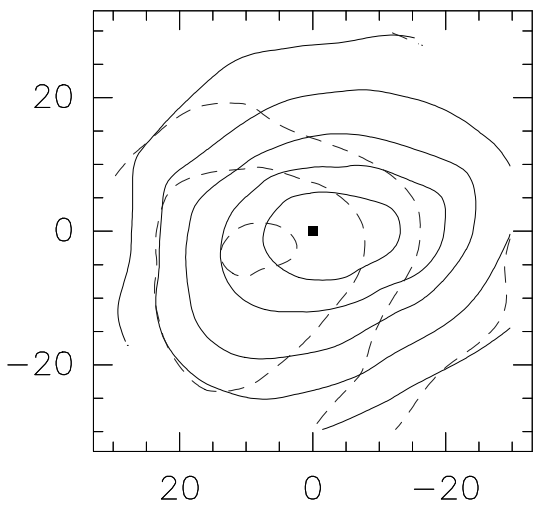
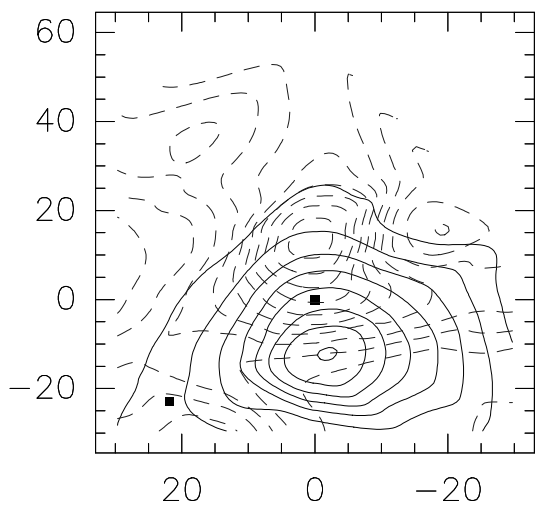
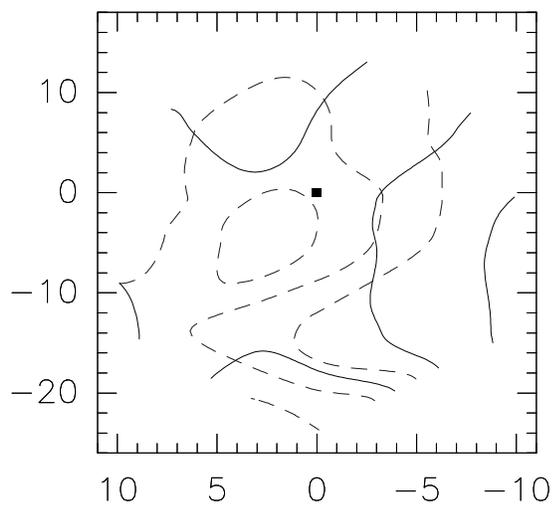
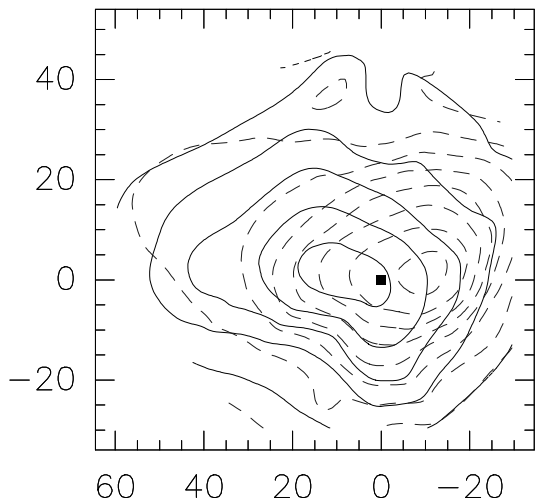
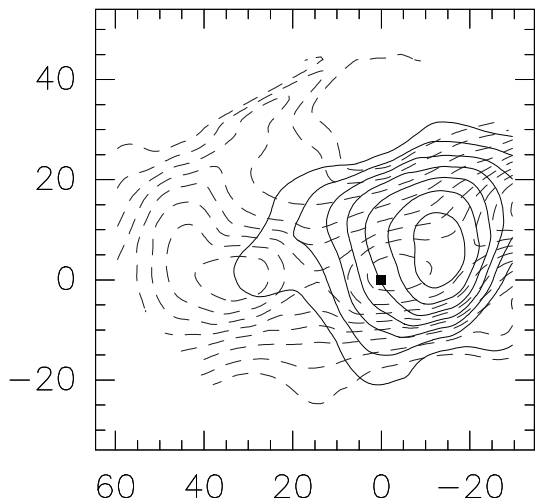
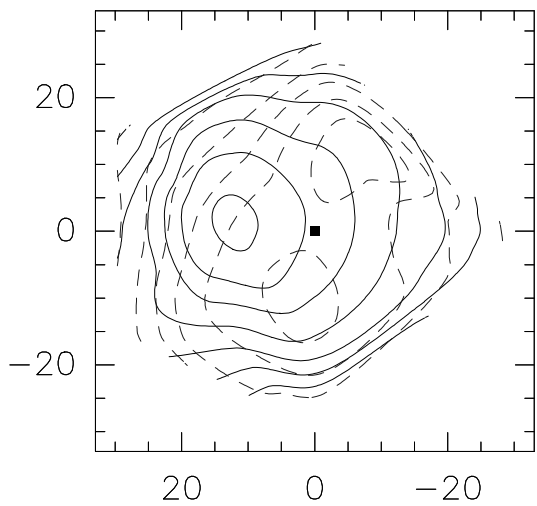
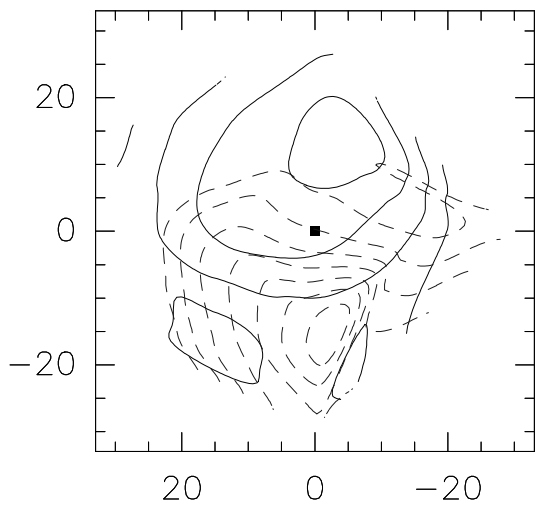
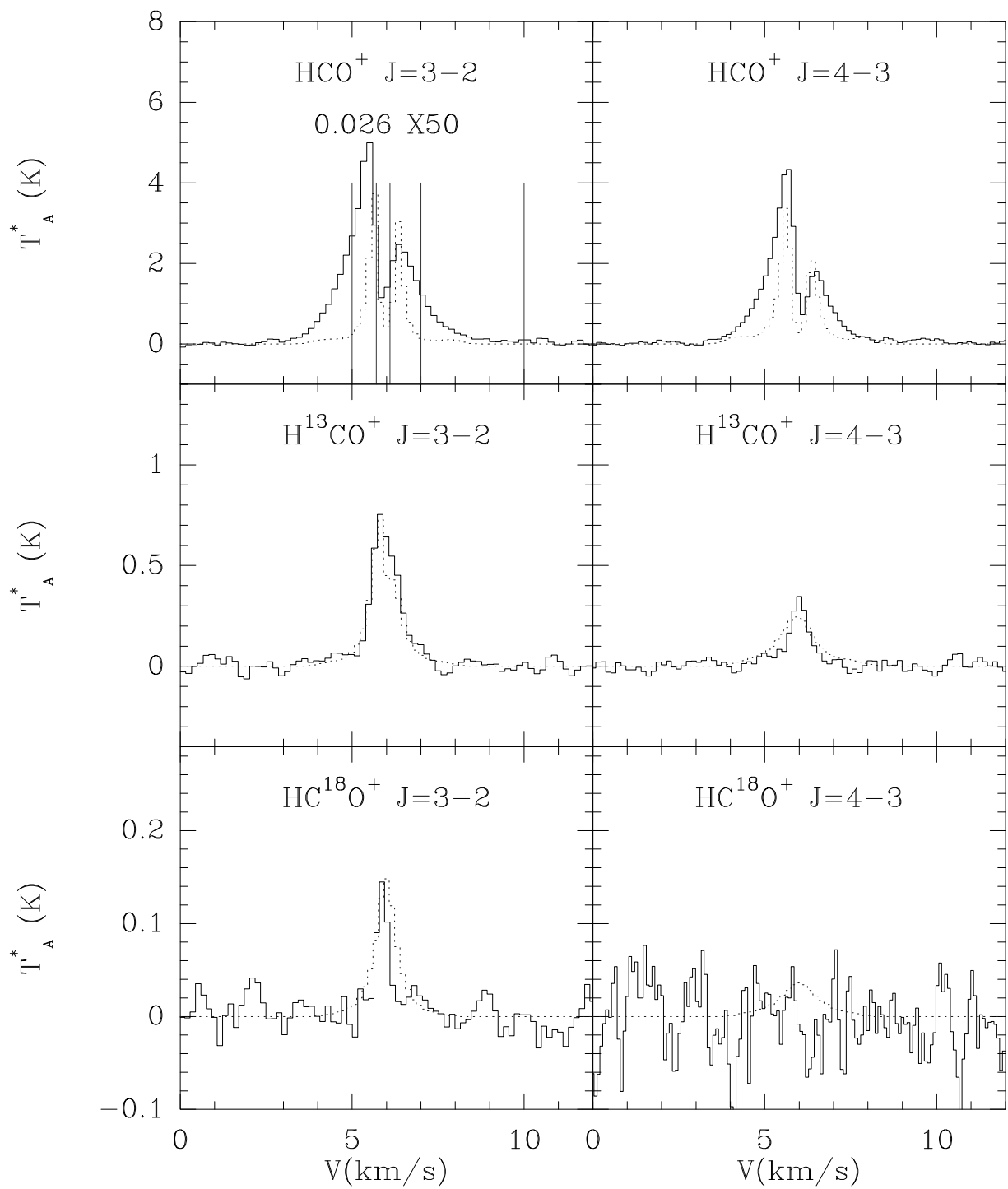
\end{document}